\documentclass{aa}   

\usepackage{graphicx}
\usepackage{txfonts}
\usepackage{pdflscape}
\usepackage[draft]{hyperref}
\usepackage{soul}

\def\L{\tilde{\Lambda}_\odot}
\def\kms{km$\,$s$^{-1}$}
\newcommand{\FeH} {[\mathrm{Fe}/\mathrm{H}]}
\defcitealias{Antoja2020}{A20}
\newcommand{\Gaia}{\textit{Gaia} }
\newcommand{\Gaiaf}{\textit{Gaia}}
\usepackage{color}

\begin{document}

   \title{Full 5D characterisation of the Sagittarius stream with \Gaiaf~DR2 RR Lyrae\thanks{Table~\ref{tab:sample} is only available in electronic form
at the CDS via anonymous ftp to cdsarc.u-strasbg.fr (130.79.128.5)
or via \href{http://cdsarc.u-strasbg.fr/viz-bin/cat/J/A+A/638/A104}{http://cdsarc.u-strasbg.fr/viz-bin/cat/J/A+A/638/A104}}}
    \titlerunning{Full 5D of the Sagittarius stream with RR Lyrae}
   \author{P. Ramos
          \inst{1}\fnmsep\thanks{email: pramos@fqa.ub.edu}
          \and
          C. Mateu\inst{2}\and T. Antoja\inst{1} \and A. Helmi\inst{3} \and A. Castro-Ginard\inst{1} \and E. Balbinot\inst{3}\and J.M. Carrasco\inst{1}}

   \institute{Dept. FQA, Institut de Ci{\`e}ncies del Cosmos (ICCUB), Universitat de Barcelona (IEEC-UB), Mart{\'i} Franqu{\`e}s 1, E-08028 Barcelona, Spain
         \and
             Departamento de Astronom{\'i}a, Instituto de F{\'i}sica, Universidad de la Rep{\'u}blica, Igu{\'a} 4225, CP 11400 Montevideo, Uruguay
        \and
             Kapteyn Astronomical Institute, University of Groningen, Landleven 12, 9747 AD Groningen, The Netherlands
             }

   \date{Received 25 February 2020; Accepted 23 April 2020}

 
  \abstract
   {The Sagittarius (Sgr) stream is one of the best tools that we currently have to estimate the mass and shape of our Galaxy. However, assigning membership and obtaining the phase-space distribution of the stars that form the tails of the stream is quite challenging. 
   }
   {Our goal is to produce a catalogue of the RR Lyrae stars of Sgr and obtain an empiric measurement of the trends along the stream in sky position, distance, and tangential velocity.}
   {We generated two initial samples from the \Gaia DR2 RR Lyrae catalogue: one selecting only the stars within $\pm$20$^\circ$ of the orbital plane of Sagittarius (Strip), and the other resulting from application of the Pole Count Map (nGC3) algorithm. We then used the model-independent, deterministic method developed in this work to remove most of the contamination by detecting and isolating the stream in distance and proper motions.
   }
   {The output is two empiric catalogues: the Strip sample (higher-completeness, lower-purity) which contains 11\,677 stars, and the nGC3 sample (higher-purity, lower-completeness) with 6\,608 stars. We characterise the changes along the stream in all the available dimensions, namely the five astrometric dimensions plus the metallicity, covering more than 2$\pi$\,rad in the sky, and obtain new estimates for the apocentres and the mean [Fe/H] of the RR Lyrae population. Also, we show the first map of the two components of the tangential velocity thanks to the combination of distances and proper motions. Finally, we detect the bifurcation in the leading arm and report no significant difference between the two branches in terms of metallicity, kinematics, or distance.}
   {We provide the largest sample of RR Lyrae candidates of Sgr, which can be used as input for a spectroscopic follow-up or as a reference for the new generation of models of the stream through the interpolators in distance and velocity that we constructed.}


   \maketitle
%

\section{Introduction}\label{sec:intro}

The Sagittarius (Sgr) dwarf galaxy \citep{Ibata1994} is the first detected and most conspicuous relic of an accretion and tidal destruction event in our Galaxy. Since its discovery \citep{Mateo1996,Totten1998}, the stream has been studied extensively using different tracers and techniques in order to map its full extent and various wraps around the Galaxy, which, combined with kinematic information,  can be used to understand the dynamics of its tidal disruption and to infer properties of the Galactic dark matter halo \citep{Law2010,Deg2013,Fardal2019}.

Compared to simpler and thinner streams, such as for example GD-1, Pal~5, or Orphan, among dozens now known in our Galaxy \citep[see][]{Grillmair2016,Mateu2018a,Shipp2018,Ibata2019}, the Sgr stream has several main characteristics that make it interesting but also challenging to observe and model. The stream is luminous and abundantly populated with stars; it is roughly planar and wraps around the Galaxy at least twice \citep{MartinezDelgado2004}; and its (observed) debris spans distances from 20 to over 100 kpc \citep{Sesar2017c}.  The extended star formation history \citep{deBoer2015} of its luminous and massive progenitor has produced complex stellar population and metallicity gradients along the stream and, because the debris is all around the sky and spans such a large distance range, it is observationally demanding to trace in a continuous manner. Despite having a high surface brightness and a stellar population that is clearly different from that of the halo, which should in principle facilitate its detection, we still lack a model that can convincingly reproduce the stream. The long-standing problem is the lack of kinematic data and the reproducibility of features such as the  bifurcation observed in both tails or the angular separation between apocentres \citep{Belokurov2006, Koposov2012,Navarrete2017,Belokurov2014b,Gibbons2016}.

One of the key elements needed to properly model the infall of Sgr is the characterisation of the spatial distribution and kinematics of its different populations throughout the sky, in a continuous and homogeneous way. \citet{Majewski2003} made the first all-sky map of the tails using 2MASS M-giants, and were followed by many later studies that obtained radial velocities for red giants, blue horizontal branch stars, and other tracers, usually in small patches along the stream \citep[e.g.][and references in \citealt{Belokurov2014b}]{Li2019}. Recently, \citet{Antoja2020}, hereafter \citetalias{Antoja2020}, used the precise astrometry of the \Gaia second data release \citep[DR2, ][]{dr2} to detect the Sgr stream from proper motion alone, without having to select a specific stellar type, and determined for the first time its proper motion along the path of the full stream. However, these latter authors did not obtain distance estimates since the parallaxes at such faint magnitudes provide little information and would require a thorough study of the astrometric systematic errors (global and local).
In any case, this would only allow distance estimates out to a few tens of kiloparsecs. Tracers for which photometric distances can be obtained are therefore more useful when studying the stream to its full extent in a comprehensive way. 

Before the publication of \Gaia DR2, \cite{Hernitschek2017} used PanSTARRS-1 (PS1; \citealt{Chambers2016}) RR Lyrae stars to measure distances to the Sgr tails and were able to trace both arms, the leading (i.e. the stars that orbit ahead of the progenitor) and the trailing (those that lag behind), all around the sky while characterising its distance and line-of-sight depth. Later, \cite{Sesar2017c} identified new Sgr features at distances over 100~kpc using the same sample. The advantage of using RR Lyrae stars is that they are excellent as standard candles, their photometric distance errors being $\sim7\%$ in the optical \citep{Mateu2012} or as good as $\sim3\%$ in the infrared \citep{Sesar2017} even without prior knowledge of their metallicity. This is an order of magnitude more precise than the $\sim 20-30\%$ errors that can be achieved with K and M giants \citep{Liu2014}, as these are far more sensitive to metallicity. For these reasons, RR Lyrae have become a standard used in studies of the halo structure \citep[for a review see Table 4 in][]{Mateu2018b} and substructure, serving to identify new streams and overdensities \citep{Vivas2001,Duffau2006,Sesar2010,Mateu2018a}; to extend and find the connection between seemingly different substructures, like the Orphan and Chenab streams \citep{Koposov2019}; and, combined with \Gaia DR2 kinematics, to provide a comprehensive 5D view of the Orphan \citep{Koposov2019} and Pal~5 streams \citep{Price-Whelan2019}. RR Lyrae, being old ($\gtrsim 10$ Gyr) and metal-poor ($\FeH\lesssim-0.5$) stars \citep{Catelan2015}, are expected to dominate the outskirts of dwarf galaxies (\citealt{Koleva2011}, and references therein) and are therefore the first to be stripped, tracing the ancient components of a stream and thus contributing to the dynamically oldest wraps.

However, an all-sky view of the kinematics of the Sgr stream with RR Lyrae is still  limited  because radial velocities are observationally demanding to obtain due to the stars' pulsations. Currently, these are only available for a few dozen RR Lyrae in selected fields along the Sgr stream \citep{Vivas2005}. 
In this work, we aim to provide a cohesive 5D view of the distance and proper motions of the Sgr stream using RR Lyrae stars. 
In Section~\ref{sec:data} we describe the input catalogue --- the RR Lyrae identified as such by the \Gaia variability pipelines complemented with the PS1 catalogue. In Section~\ref{sec:methods} we derive distances for these RR Lyrae stars and describe our selection method  based on sky coordinates, proper motions, and distances, and use no prior information. In Section~\ref{sec:properties} we present our results providing an entirely empirical characterisation of the stream including for example its proper motion, Galactocentric distance, tangential velocities, and the bifurcation. Finally, we present our conclusions in Section~\ref{sec:conclusions}.


\section{RR Lyrae and \Gaia sample}\label{sec:data}

The second data release \citep{dr2} of the \Gaia mission \citep{gaiamission} has provided not only positions, magnitudes, and proper motions for more than a billion stars, but also a vast catalogue of variable sources \citep[hereafter VC, see ][]{Holl2018}. For a detailed description of the detection, classification, and post-processing pipelines we refer the reader to \cite{Eyer2017} and \citet{Rimoldini2019}. Among those detected variable sources, there are 140\,784 RR Lyrae that have at least 12 good $G$-band transits which have been validated by the Specific Objects Study (SOS) pipeline \citep{Clementini2019}. 
We have further increased the number to 228\,904 sources by including those classified as RR Lyrae by \cite{Holl2018} and \cite{Rimoldini2019}, but not processed by SOS due to the small number of observations available in DR2. We then apply the filter recommended in \cite{Rimoldini2019} to remove obvious contaminants: \begin{tt}phot\_bp\_rp\_excess\_factor > 2 OR NULL\end{tt}, leaving 175\,164 RR Lyrae. Finally, we include 11\,318 stars identified as \emph{bona fide} RR Lyrae in the  PS1 catalogue \citep{Sesar2017}, but not classified as such by the \Gaia pipelines. We identify these stars in the DR2 \begin{tt}gaia\_source\end{tt} table  by cross-matching their positions on the sky with a $1"$ tolerance, and retrieve their astrometric data.

After removing the stars with no proper motions we obtain a list of 182\,495  stars, 122\,745 of which have been processed by the SOS pipeline. For some of them, the SOS team has derived photometric metallicities and absorption in the G band. In particular, 39\,129 of them have both quantities simultaneously,  which we use in Sect.~\ref{sec:dist} to obtain a first measurement of the metallicity distribution along the stream. As stated in \cite{Holl2018}, the \Gaia catalogue of variable sources is not meant to be complete and we do not expect the completeness to be above 80\% when taking only sources with more than 12 field-of-view (FoV) transits (see also Mateu et al., in prep.).


\section{Selection of RR Lyrae in the Sagittarius stream}\label{sec:methods} 

Our aim is to detect the RR Lyrae of the  Sgr stream from scratch to produce a characterisation that is as empirical as possible. To do so, we select 
two sets of initial candidates from the list of stars with proper motions presented in the previous section: the Strip and the nGC3 selections (Sect.~\ref{sec:initial_sample}), later pruned to filter out contaminants based on distance (Sect.~\ref{sec:members}) and kinematic information (Sect.~\ref{sec:pmra} and \ref{sec:pmdec}). We use these two different selections since they offer different advantages as we shall see below. In Table~\ref{tab:counts} we show the number of stars remaining after each selection step.

\begin{table}
\caption{Number of stars in each sample at each step of the selection process.}\label{tab:counts}
\centering
\begin{tabular}{lcc}
\hline\hline
Stage & nGC3 & Strip \\
\hline
Initial sample & 13\,004 & 76\,872 \\
Distance selection & 7\,953 & 18\,045 \\
$\mu_{\alpha*}$ selection & 6\,797 & 12\,583 \\
$\mu_{\delta}$ selection& 6\,608 & 11\,677 \\
\hline
\hline
\end{tabular}
\end{table}

\subsection{Initial selections: Strip and nGC3}\label{sec:initial_sample}

\begin{figure}
    \centering
    \includegraphics[width=1\linewidth]{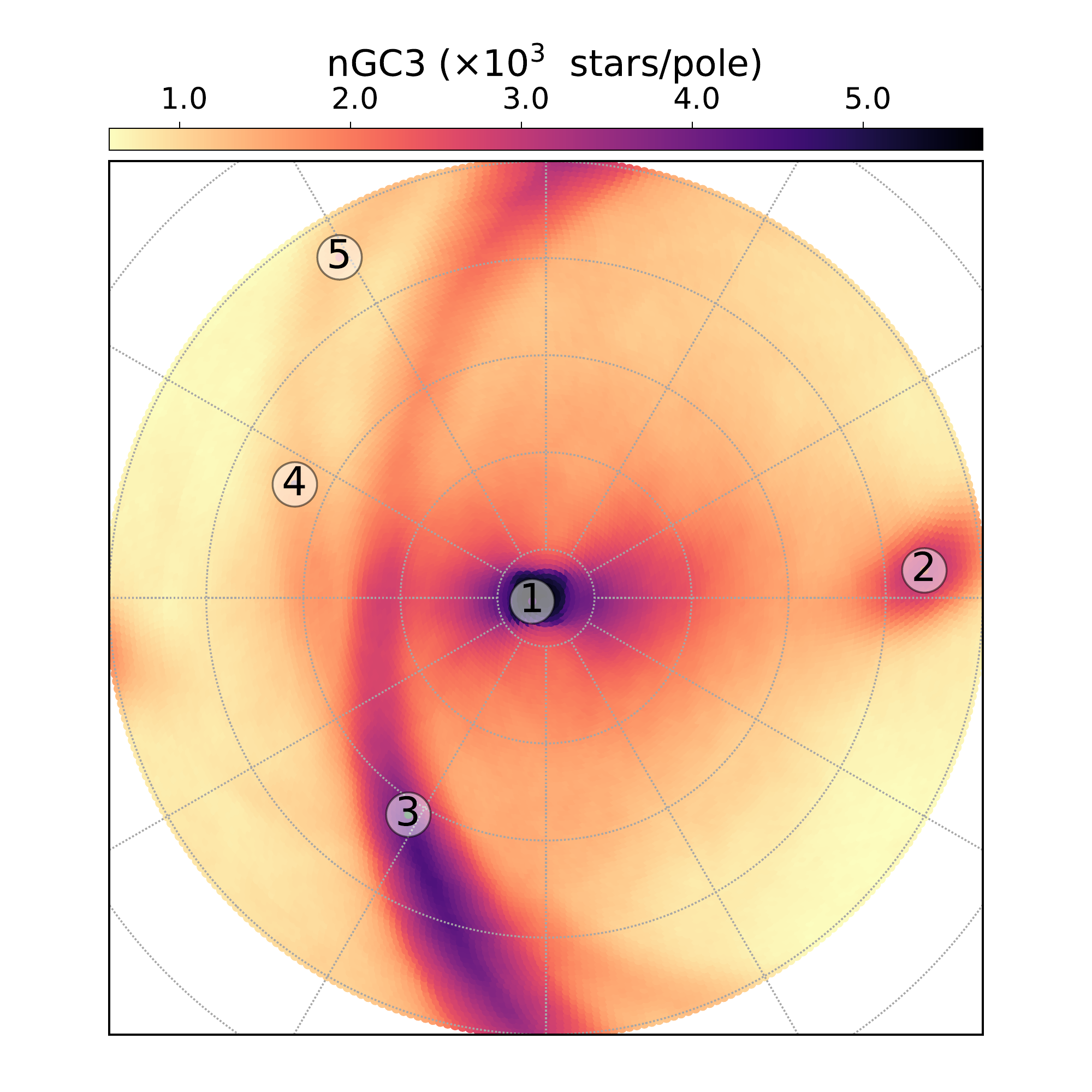}
    \caption{Pole Count Map of the RR Lyrae catalogue described in Sect.~\ref{sec:data}. The Galactic disc produces the large concentration at (1), while the peaks (3),(4), and (5) are related to the Large and Small Magellanic clouds. The prominent signal on the right, (2), is caused by the Sgr stream.}
    \label{fig:PMC}
\end{figure}

\paragraph{\textbf{Strip selection}. A high-completeness, low-purity sample:}
A first straightforward list of candidates can be obtained by selecting stars within 20º of the orbital plane defined by the pole  (l,b)=(273$\degr$.8; 13$\degr$.5) from \cite{Majewski2003}. This is exactly equivalent to selecting stars with  |$\tilde{B}_{\odot}$|$\,$<$\,$20$\degr$ in the Sgr coordinate frame\footnote{In this work, we use the convention by \cite{Belokurov2014b}.}, a spherical heliocentric frame rotated to have the plane of the stream at $\tilde{B}_{\odot}=0\degr$ and the Sgr dwarf remnant at $\tilde{\Lambda}_{\odot}=0\degr$. The resulting selection contains 76\,872 RR Lyrae (41\% of the full catalogue). 

This selection is expected to be highly complete, limited only by the completeness of the input catalogue itself, as no kinematic or additional information (e.g. metallicity) is required. However, precisely because no information other than sky position is used, a large fraction of contaminants is expected, mostly due to the thick disc and halo. This contamination will be reduced significantly after the following selection steps.

\paragraph{\textbf{nGC3 selection}. A low-completeness, high-purity sample:}

Great circle cell count methods grouped in the xGC3 family \citep{Johnston1996,Mateu2011,Mateu2017} are aimed at the detection of groups of stars that populate an orbital plane different from that of the Milky Way disc. In its most basic implementation, this approach relies on the fact that the stars in a stream lie approximately on a plane through the centre of symmetry of the Galactic potential. Geometrically, this implies that the stream is confined to a great circle uniquely defined by the normal vector (pole) of the orbital plane. Therefore, if we grid the sky\footnote{For streams that are too close to the Sun, projecting their orbit onto the Celestial sphere might introduce projection effects, i.e. the Galactic parallax becomes non-negligible.} and, for each cell, count the number of stars in a wafer of a certain width perpendicular to it, we should detect an overdensity whenever we align this wafer with the orbital plane of a stream. The result is a pole count map (PCM), which is a density map of the number of stars associated to each pole and, hence, great circle.

The nGC3 method introduced in \citet{Abedi2014}  also requires proper motion information in the pole counting. In this method, the velocities of the stars are required to lie in the same great circle as their positions, within a certain tolerance. This helps to clean the sample of contaminant stars that lie in a given great circle band by chance. We cannot take this approach in the heliocentric frame because then the peculiar velocity of the Sun would dominate the PCM. Instead, the nGC3 pole counts are made in the galactocentric reference frame, for which we assume a distance from the Sun to the Galactic Centre (GC) of $R_\odot$=8.0 kpc \citep{Camarillo2018a}, a velocity of the Sun with respect to the local standard of rest (LSR)  [$U_\odot$,$V_\odot$,$W_\odot$] = [11.10, 12.24, 7.25] \kms{} \citep{Schonrich2010}, and $V_{LSR}$ = 240 \kms{} \citep{Reid2014}. Obviously, we also require a distance to each star which implies assuming a metallicity (see Sect.~\ref{sec:dist}). We chose as a first value [Fe/H]$\,$=$\,$-1.7 dex. We find that the PCM does not change significantly within the range of metallicities typical for RR Lyrae in the halo: -1.5 dex $\leq$ [Fe/H] $\leq$ -1.7 dex. 

Figure \ref{fig:PMC} shows the resulting PCM for the RR Lyrae, using a tolerance of 5$^{\circ}$ for both positional and velocity vectors. The method also returns a list of peaks, that is, local maxima, ordered by the number of counts contained within the corresponding pole. As expected, we observe a peak (label 1) at the centre caused by the disc. The second most dominant peak (label 2) is that produced by the Sgr stream. We also see signatures produced by the Large (label 3) and Small (labels 4 and 5) Magellanic clouds which, being concentrated on the sky, make trails instead of compact overdensities. 

The nGC3 selection is made by extracting the RR Lyrae stars associated to the cells, or poles, around peak 2 within the 25th percentile of the maximum counts in that peak. This results in a preliminary list of candidates of 13\,004 (7\% of the full RR Lyrae catalogue). This sample is expected to be of lower completeness than the Strip selection because the method can introduce correlations between the spatial position and the proper motions and, more importantly, a dependence on the observational errors which can translate into kinematical biases. In particular, the method by construction selects against stars with significant motions perpendicular to the plane of the  stream, making it unsuitable for the study of aspects such as velocity dispersion profiles, as we discuss in Sect.~\ref{sec:vel_disp}. By contrast, the nGC3 sample is expected to be of higher purity, and therefore better suited for purposes such as the selection of targets for spectroscopic follow-up. 

Having a first list of candidates, we next look for trends in distances and proper motions to refine our selection. For the remainder of this section only, to avoid repetition, we use the nGC3 sample  to show how our selection methodology works. This sample, by construction, is less contaminated than the Strip sample and allows us to  more clearly illustrate the way we separate the candidate stars from the contamination.  Wherever important differences with the Strip sample arise they are discussed in the text. The plots corresponding to the Strip selection can be found in Appendix~\ref{app:strip}. 


\subsection{Distance determination}\label{sec:dist}

To determine the distances we begin by using the linear relation for the \Gaia $G$ band given in \cite{Muraveva2018}:

\begin{equation}\label{eq:LvsZ}
    M_G = 0.32^{+0.04}_{-0.04}\FeH + 1.11^{+0.06}_{-0.06}.
\end{equation}
\noindent This relation returns the absolute magnitude of the RR Lyrae star given its metallicity in the \cite{Zinn1984} scale (hereafter, ZW84). However, we note that the metallicity given by DPAC is calculated from the Fourier parameters of the light curve 
\citep{Clementini2019}. In particular, in the case of RR Lyrae of type RRab, the metallicity is calibrated with the prescription of \cite{Nemec2013} which, for the range of periods and Fourier parameters of our sample (c.f. their Fig. 12), closely matches the \cite{Jurcsik1996} metallicity scale.
As noted in \cite{Gratton2004} and \cite{DiFabrizio2005}, there is a systematic difference of +0.3 dex between the metallicities derived from photometry in the \citet{Jurcsik1996} scale and those calculated with spectroscopy\footnote{The offset was calculated with respect to the \cite{Harris1996} spectroscopic scale which has almost no shift compared to the ZW84 scale used in this work \citep{Gratton2004}.}. Therefore, for our particular case, we subtract 0.3 dex from each RRab to convert the metallicities from DPAC into the ZW84 scale. This is not the case for the stars of type RRc, as their metallicity is given in a scale already similar enough to ZW84 \citep{Nemec2013,Clementini2019}.

Even though we have a way to compute the absolute magnitude for our stars, the metallicity is only available for a small fraction of them ($\sim$35\%, see \citealt{Clementini2019}). Additionally, the Sgr stream presents a gradient in metallicity (e.g. \citealt{Zhang2017,Hayes2019}). Our approach is to first  test whether there is a metallicity gradient for the RR Lyrae stars with measured metallicity. This is important because it could introduce a distance bias along $\L$. However, we find that there is not (see below), and we assign an average metallicity to the rest of RR Lyrae.

To calculate the distances, we use the well-known relation for the apparent magnitude of a star,
\begin{equation}\label{eq:dist}
    d \,[pc] = 10^{(m_G-M_G-A_G)/5+1},
\end{equation}
for all stars with reported metallicity and absorption in the nGC3 sample. In particular, we choose a sample of 884 stars between 20 kpc and 50 kpc from the Sun, i.e. preferentially selecting stars belonging to the stream (c.f. Fig.~4 of \citealt{Hernitschek2017}), all of which happen to be RRab according to the classification by SOS. 

\begin{figure}
    \centering
    \includegraphics[width=1\linewidth]{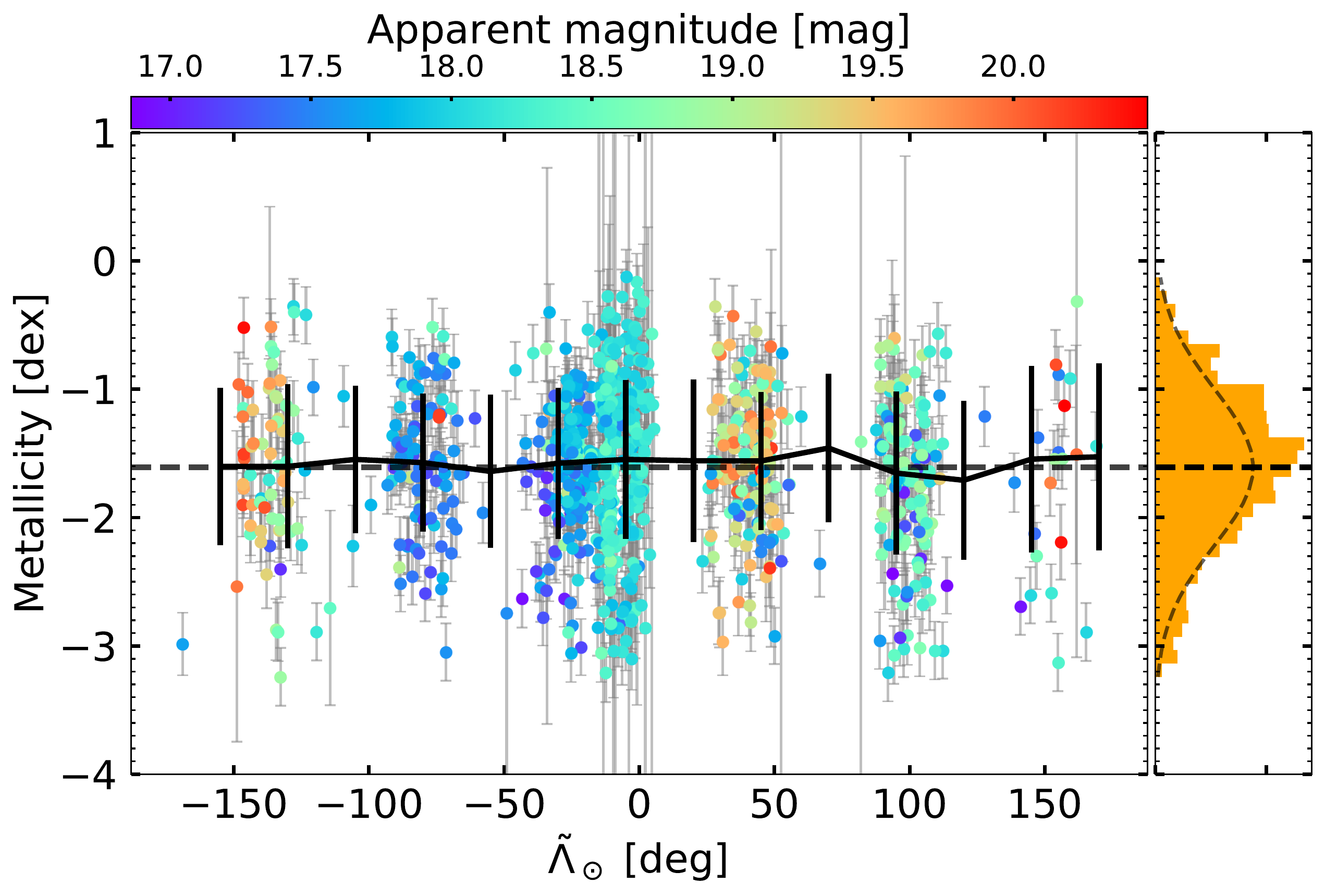}
    \caption{Metallicity dependence on $\L$ for the stars selected by the \textit{nGC3} method that have a measured metallicity and absorption. Only those stars between 20 kpc and 50 kpc are shown (884 RRab), coloured according to their apparent magnitude. The dashed line corresponds to the mean metallicity (-1.61 dex) whereas the solid line is the running median and 1$\sigma$ interval within bins of 50$^\circ$ every 25$^\circ$. On the right, we include the metallicity distribution and a Gaussian fit to the data: $\mu$\,=\,-1.61 dex and $\sigma$\,=\,0.61 dex.}
    \label{fig:FeHvsL}
\end{figure}

In Fig. \ref{fig:FeHvsL} we study the dependence of the metallicity on $\L$. The gradient we measure is  (-1.5$\pm$0.4)$\times 10^{-3}$\,dex\,deg$^{-1}$ and (0.9$\pm$3.7)$\times 10^{-4}$\,dex\,deg$^{-1}$ for the leading and the trailing arms, respectively. While the latter is clearly compatible with zero, the former shows some sign of a gradient. Still, this gradient can be heavily influenced by the contamination, and the variations in metallicity that this would imply (0.15\,dex every 100$^\circ$) are smaller than the typical photometric metallicity precision ($\sim$0.2\,dex). Therefore, we recover a mean metallicity $\overline{\FeH}_{ZW}$=-1.61 dex with no trace of a gradient, in good agreement with previous studies of the RR Lyrae population of Sgr \citep{Cseresnjes2001}. As for the statistical uncertainty on the mean, we find $\frac{\sigma_{\FeH}}{\sqrt{N}}$ = 0.02 dex. After repeating the measurement for the Strip sample, we obtain the same result but with a smaller uncertainty ($\frac{\sigma_{\FeH}}{\sqrt{N}}$ = 0.01 dex). We re-visit this determination in Sect.~\ref{sec:metallicity}, after having selected a reliable sample of Sgr stars, and we obtain the same value within the uncertainties.  

Given that there is no significant gradient, we impose the mean metallicity on all the candidate stars. We note that the distribution of apparent magnitude of the RR Lyrae type C is the same as that of the RRab. Assuming that there is no segregation between both types inside the stream, their distance distribution at any given $\L$ should then be the same. Therefore, we can use the same metallicity for both as long as we apply the same calibration, i.e. Eq. \ref{eq:LvsZ}. 

In addition, to obtain a distance estimation for each star we need the absorption in the $G$ band. For that, we use the \citet{Schlegel1998} dust maps with the \citet{Schlafly2011} re-calibration and the relations that allow us to translate reddening to absorption (Appendix~\ref{app:reddening}). Having done that, we apply Eqs.~\ref{eq:LvsZ} and \ref{eq:dist} to the whole list of candidates and propagate the uncertainties in metallicity, apparent magnitude, coefficients of Eq.~\ref{eq:LvsZ}, and absorption\footnote{We find that, when propagating the errors in colours to the transformations from $A_V$ to $A_G$ by Monte Carlo sampling, the uncertainty in absorption can be taken to be 0.02 dex.} using the Jacobian of the transformation. 

Apart from the statistical uncertainties, we also have a systematic source of uncertainty: our metallicity zero point. As mentioned above, we subtracted 0.3\,dex from the metallicities reported by SOS to obtain values in the ZW84 scale. Nevertheless, if that is not the right offset we would be biasing all distances by a fixed proportion:
\begin{equation}\label{eq:systematic}
    \frac{D(\Delta Z)}{D_0} = 10^{-\frac{0.32}{5}\Delta Z},
\end{equation}
\noindent
where $\Delta Z$ is the difference between the zero point that we assume and the true one. For instance, having no zero point in reality ($\Delta Z$=\,+0.3\,dex) and having a zero point that is double what we assumed ($\Delta Z$=\,-0.3\,dex), respectively, and for a distance of 25\,kpc, we would obtain 23.92\,kpc (-4.3\%) and 26.13\,kpc (+4.5\%). We expect the true zero point to be closer to our value and therefore the systematic error to be much smaller than $4\%$.

\subsection{Distance selection}\label{sec:members}
\begin{figure}
    \centering
    \includegraphics[width=\linewidth]{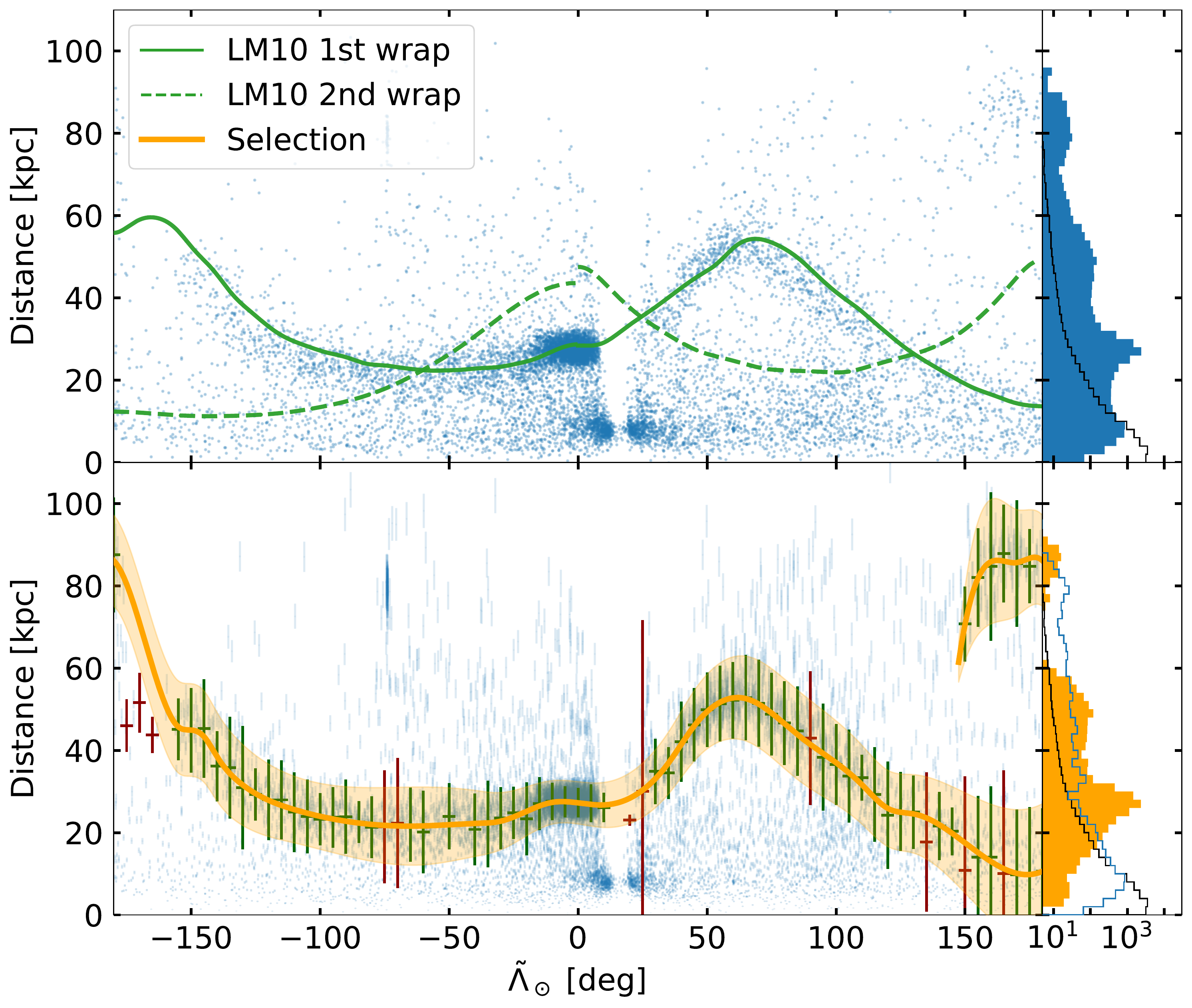}
    \caption{Distance as a function of $\L$ for the stars in the Sgr stream. Top: All candidates selected by the \textit{nGC3} method. The green lines represent the predictions of the LM10 model for the first (solid) and second (dashed) wrap. The blue bars on the right are the histogram of the same stars, while the black steps correspond to the predictions from the mock catalogue (see text). Bottom: Same stars, now with their associated uncertainties, with green (red) error bars representing the parameters extracted from the good (bad) fit to the kernel associated with Sgr (see text). We also show in orange the interpolation of the centres (dashed line) and the two-sigma interval (shaded area). The histogram on the right  now shows the selected stars in orange, the remaining ones in blue, and the black one is the same as above.}
    \label{fig:LvsDist}
\end{figure}

In the top panel of Fig. \ref{fig:LvsDist} we present the resulting distribution of distances with $\L$. A conspicuous pattern emerges as a result of selecting sources belonging to the Sgr orbital plane. Nevertheless, there is still a noticeable fraction of contamination even after applying the nGC3 PCM technique: thick disc, halo, and even the Sculptur dwarf Spheroidal at $\L\sim$-75$\degr$ (the latter is easily removed). Continuing with the aim of re-discovering the stream from scratch, we apply a method to select probable members that requires almost no assumptions and that we call Tailored Gaussian Mixture (TGM).

The idea of the TGM is to select the pattern formed by the Sgr stars with $\L$ isolating it from both the foreground and the background contamination. For a given bin of $\L$, we process the histogram of heliocentric distances and identify the different components (Sgr and contamination). A Gaussian is then fitted to each component to obtain their widths. The algorithm starts at the dwarf and uses it as an anchor since it is the dominant component at $\L$\,=\,0$^\circ$. The component corresponding to Sgr in the contiguous bins is assigned by finding the one with the highest continuity in heliocentric distance. 
When none of the components found are close enough to the one in the previous bin, we skip the bin and try with the next one. The details of the method can be found in Appendix~\ref{app:TGM}.

With this algorithm we obtain a deterministic measurement of the centre of the pattern as well as a first estimation of its depth. 
The result for the nGC3 sample is shown in the bottom panel of Fig. \ref{fig:LvsDist} as error bars located at the centre of each bin in $\L$. The vertical extent of the error bars represents the 2$\sigma$ interval of the dispersion recovered by the fit and the horizontal extent represents the size of the bin (barely visible).
We note that using the same metallicity for all the stars increases the contamination because the foreground (thick disc) merges with the signal of the stream.  Since the RR Lyrae in the thick disc tend to be more metal rich than those in Sgr, assigning an incorrectly smaller  metallicity to these stars translates into larger distances and thus pushes its distribution towards that of the stream, hampering the separation. Finally,  by visual inspection we find that in some of the bins the Gaussian fit has not converged to a good solution. The two main causes are i) too few sources and ii) the components are too close to be resolved, resulting in an overestimated width. Those few cases (plotted in red in the bottom panel of Fig. \ref{fig:LvsDist}) have been removed.

We finish by creating two splines with the good solutions (green error bars). The first passes through the centres and represents the change in the distance with $\L$ (orange line). We use the second one to trace the changes in depth of the stream and produce a smooth $\sim$2$\sigma$ confidence interval, as represented by the shaded orange area. We also show the smoothed median distance predicted by the \cite{Law2010} model (hereafter, LM10\footnote{Downloaded from http://faculty.virginia.edu/srm4n/Sgr/}). These solid (first wrap) and dashed (second wrap) green lines can be seen to follow our orange track in some part of the stream. However, there are important differences which we comment on in Sect.~\ref{sec:properties}. On the other hand, our results are in very good agreement with the recent determination of the pattern of Sgr obtained with RR Lyrae by \cite{Hernitschek2017} (see their Fig.~4).

The candidates are selected if they fall inside the orange shaded area, resulting in a set of 7\,953 stars (61\% of the initial nGC3 sample). In the case of the Strip sample (Fig. \ref{fig:LvsDist_sky}) the output of the TGM results in a selection of\,18 045 stars (23\%). Both samples contain field stars that require additional filters to be discarded, especially at $\L\geq$120º where the Sgr stars get closer to the Sun and merge with the Virgo overdensity \citep{Juric2008}. In that region we also have the presence of Sgr stars at a distance of $\sim$90\,kpc, which correspond to the apocentre of the trailing arm.

We also include a comparison with a mock catalogue. We query the mock \Gaia catalogue by \citet{Rybizki2018} in HEALPix bins of level 4 ($\sim$14 square degrees) along the orbital plane of Sgr selecting 2\,000 stars for each patch in the sky. We then compute the true distances by simply inverting the true parallaxes and produce the heliocentric distance histograms shown in black on the right of Fig. \ref{fig:LvsDist} (normalised to the number of stars in the nGC3 sample). The mock distribution presents a smooth exponential profile with no sign of bumps, in contrast with our data (blue filled histogram). With the mock as a reference, we gain an idea of the contamination that we expect at each portion of the stream, with the parts near pericentre being  the most susceptible. This corresponds to the beginning of the trailing tail and the far end of the leading one ($\L$>\,120$^\circ$),  this latter being particularly contaminated. After performing our selection (orange histogram at the bottom right panel of Fig.\ref{fig:LvsDist}) we see that what we have discarded, that is, what we consider contamination (blue line), resembles the mock distribution, except for two details: i) at short distances, the effect mentioned earlier of using an incorrect metallicity for the thick disc modifies the shape of the distribution and ii) the absolute value of counts at large distances is not the same. The latter could be due either to a problem of normalisation between the thick disc and halo components of the mock or simply to the fact that the parameters of the model do not reproduce the observations accurately enough.

\subsection{Proper motion selection: right ascension}\label{sec:pmra}

We continue the selection of RR Lyrae candidates based now on their proper motion trends throughout the Sgr stream. 
In order to avoid overlap between the different wraps, we take those corresponding to the apocentre of the trailing arm and shift their location by -360$^\circ$. Therefore, from here onwards the plots reach beyond -180$^\circ$. 

\begin{figure}
    \centering
    \includegraphics[width=\linewidth]{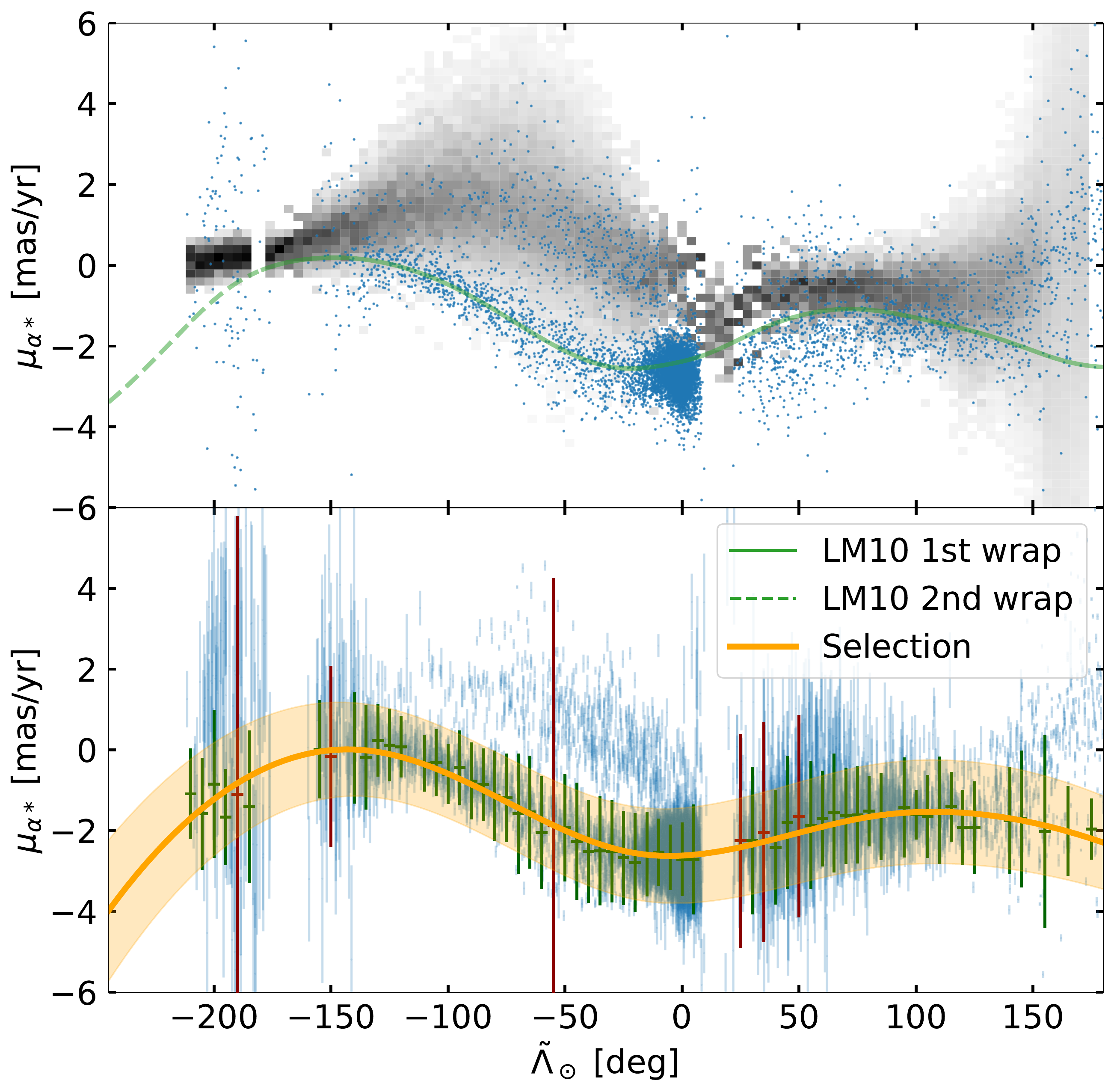}
    \caption{$\L$ against proper motion in right ascension of the Sgr stream. Top: All stars selected by distance superimposed on the histogram obtained from the mock catalogue (see text). Bottom: Result of the TGM (green and red error bars) and the area that we use to further filter out the contamination (orange). Here we also show the prediction of the LM10 model with green lines.
    The stars at the trailing apocentre have been moved to the left of the plot by subtracting -360$^\circ$.}
    \label{fig:pmra}
\end{figure}

Taking the stars already selected according to their heliocentric distance, we now plot in Fig.\ref{fig:pmra} the observed proper motion in right ascension as a function of $\L$ (blue dots). We observe a modulation of the dispersion which is dominated by the correlation between the uncertainty and the distance. This is especially noticeable in the Strip sample (Fig. \ref{fig:pmra_sky}), where the dispersion near the apocentres is significantly larger. Also, there are at least two clearly identifiable patterns, especially at the beginning of the trailing and the leading arms.

To understand the effects of the contamination, we resort once more to the mock MW catalogue described in Sect.~\ref{sec:members}. We use the orange area obtained in Sect.~\ref{sec:dist} to filter the particles in the model by distance, add the errors in proper motion provided with the mock catalogue and finally create the 2D histogram shown in the top panel of Fig. \ref{fig:pmra}. As can be seen, it follows quite closely the trail of stars at higher $\mu_{\alpha*}$. We note that these mock sources are all labelled as halo stars, except for the small range of $\L$ where the stream passes behind the GC, where it is dominated by thick disc stars with a small fraction of thin disc stars.

As done for the distance, we run our TGM with the parameters listed in Appendix \ref{app:TGM} and skip any bin that contains less than five stars because we have fewer left after the previous selection. The results are shown in the bottom panel of Fig. \ref{fig:pmra}, where we can already see that the prediction of the LM10 model shown in the upper panel (green line) is similar to the trend we detect (see Sect~\ref{sec:pm} for a discussion on the residuals). Again, after selecting only the stars within the orange shaded area, we are left with 6\,797 RR Lyrae (52 \%) stars in the nGC3 sample. In the case of the Strip sample, we find that we cannot recover the same trend beyond $\L\geq$120$^\circ$ due to the predominance of the contamination. We tried different parameter combinations without any success. Finally, to avoid leaving that region void, we  only select stars around the track of the LM10 model beyond $\L=$120$^\circ$, choosing the ones with -3.5 mas$\,$yr$^{-1} \leq \mu_{\alpha*} \leq$ -0.5 mas$\,$yr$^{-1}$ (orange box in Fig. \ref{fig:pmra_sky}). This results in a list of 12\,583 stars (16 \%).

\subsection{Proper motion selection: declination}\label{sec:pmdec}
\begin{figure}
    \centering
    \includegraphics[width=\linewidth]{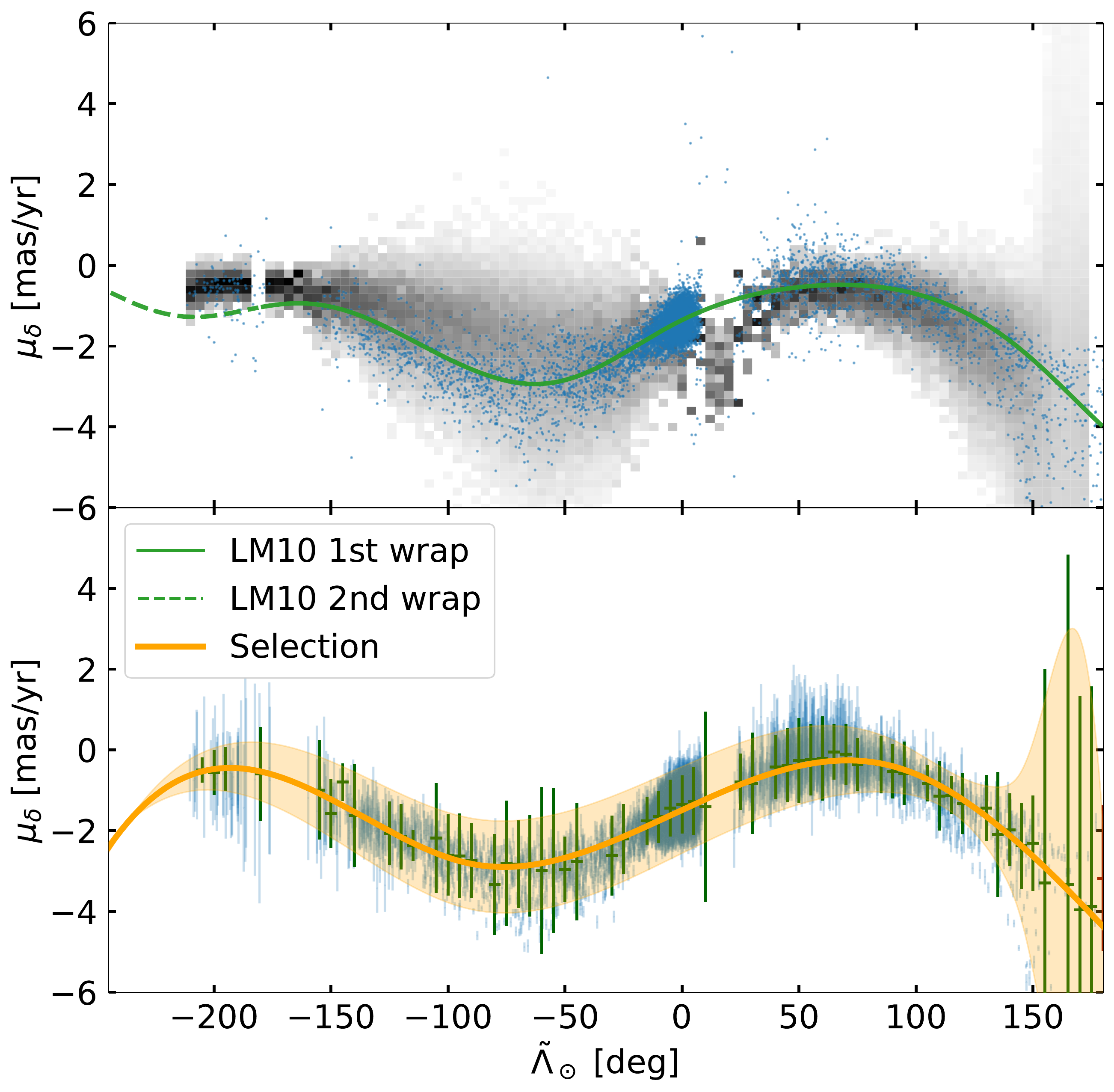}
    \caption{Same as Fig. \ref{fig:pmra} but with proper motion in declination. Now we only show the stars selected by distance and $\mu_{\alpha*}$.}
    \label{fig:pmdec}
\end{figure}

We apply the TGM now to the proper motion in declination of the remaining stars with the same parameters as in Sect.~\ref{sec:pmra} and recover a single component all along the stream, as can be seen in Fig. \ref{fig:pmdec}. We note that the halo stars from the mock catalogue present a proper motion that follows  that of the Sgr stream closely but not exactly. This is likely due to the fact that $\mu_{\delta}$ is roughly aligned with the motion perpendicular to the stream\footnote{The ICRS frame and the Sgr plane defined by \cite{Majewski2003} have similar poles.}, and because the dominant contribution in this direction is the Solar reflex, it is natural that the stream cannot be cleanly separated from the field. In this step,  we have therefore performed the equivalent to a two-sigma clipping.

\noindent
After applying these selections we obtain our final list of candidates with 6\,608 stars for the nGC3 sample and 11\,677 for the Strip. The two catalogues are available as one table at CDS (see Appendix~\ref{app:interpolators}), with a flag that describes whether the star was selected using one method, the other, or both.

\section{Results}\label{sec:properties}

Now that we have assembled a reliable list of stars belonging to the Sgr stream, we proceed on to examine its main characteristics. In this section, in contrast with the previous, we focus on the Strip sample because, as we discuss below, the nGC3 presents a kinematical bias induced by the PCM method. The corresponding figures and derived quantities for the nGC3 sample can be found in Appendix~\ref{app:nGC3_summary}.

\subsection{The missing dimensions: proper motions}\label{sec:pm}

\begin{figure}
    \centering
    \includegraphics[width=\linewidth]{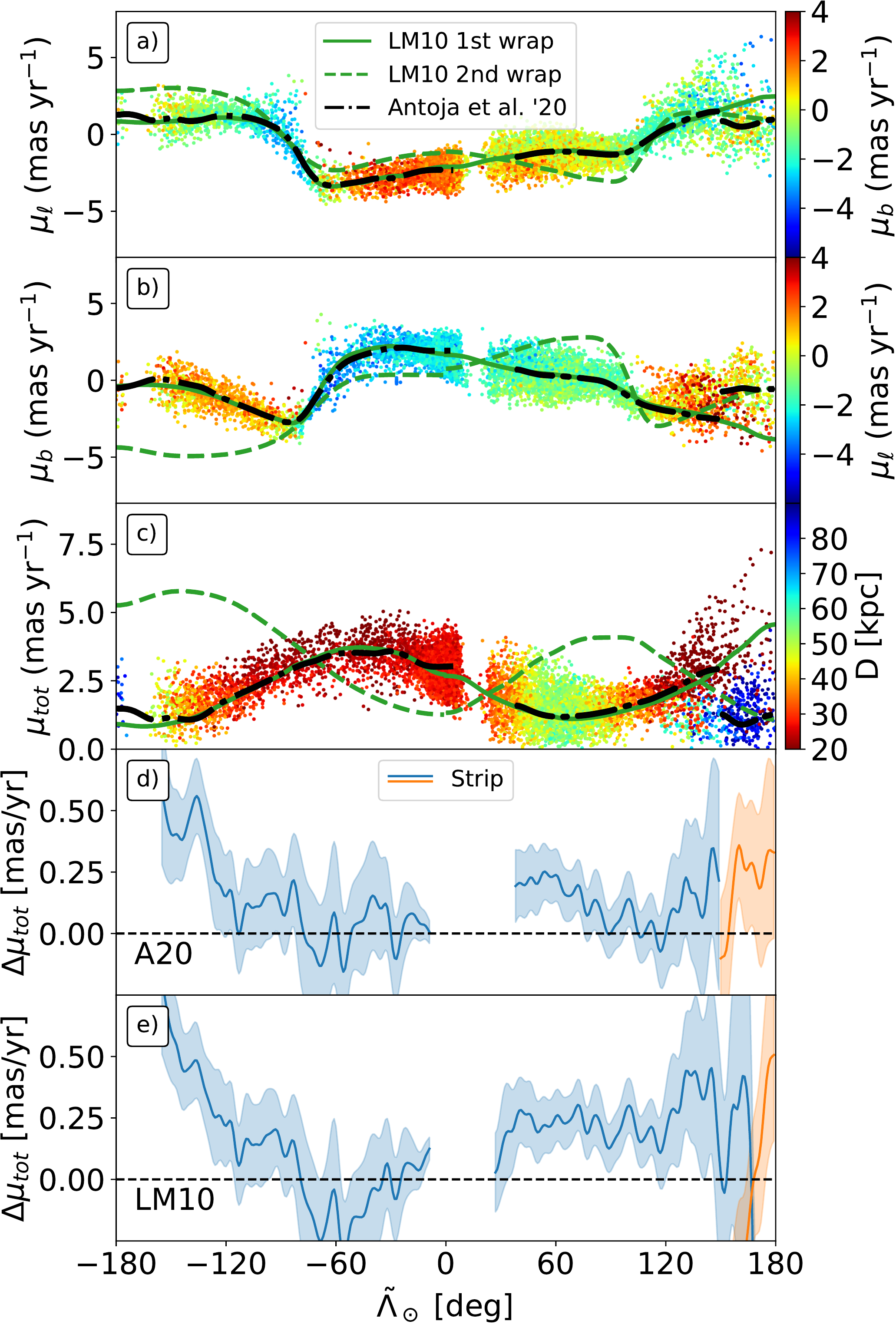}
    \caption{Proper motion trends with $\L$ of the candidate RR Lyrae of the Sgr stream in the Strip sample. a) Proper motion in Galactic longitude coloured according to proper motion in Galactic latitude. b) Same as above but substituting longitude with latitude. c) Total proper motion coloured by distance. In all these panels, we show the predictions from the LM10 model with solid (first wrap) and dashed (second wrap) green lines, as well as the smoothed median from \citetalias{Antoja2020} (dashed-dotted black line). Panel (d) contains the residuals of the smoothed median of the total proper motion (blue - first wrap, orange - second wrap) minus that of \citetalias{Antoja2020}. Panel (e), similarly, shows the residuals with respect to the LM10 model.}
    \label{fig:ProperMotion}
\end{figure}

First, we want to determine the proper motion of the Sgr stream and examine the two samples looking for the presence of any kinematic biases. We do so by comparing the trends in proper motion of our stars with those obtained by \citetalias{Antoja2020} and the predictions of LM10. In Fig.~ \ref{fig:ProperMotion} (Fig.~\ref{fig:ProperMotion_ngc3}) we present the proper motions in Galactic coordinates as a function of $\L$ for the Strip (nGC3) sample. The colours of the panels (a), $\mu_l$, and (b), $\mu_b$, as indicated by the colour bars, are the same as in Fig.3a of \citetalias{Antoja2020} for ease of comparison. In panel (c) on the other hand, we show the total proper motion coloured according to the distance derived for each RR Lyrae star. 
In all three panels, we have added the smooth median obtained from \citetalias{Antoja2020} (black) and the predictions of the LM10 model (green). We note that the farthest stars (blue points in panel c) have approximately the proper motion that the model predicts for the second wrap, confirming that we are indeed detecting the apocentre of the trailing tail. 

In panel (d) we plot the difference between the median of our sample and that of \citetalias{Antoja2020}, both smoothed with a Gaussian filter,  and the associated 1$\sigma$ intervals. While the discrepancy is within 2$\sigma$ in most parts of the stream, we note that the residuals start to grow again at $\L$>120\degr, which is probably related to the degree of contamination being higher in this region due to our selection (see Sect.~\ref{sec:pmra}). In the trailing tail, the differences are within 1$\sigma$ until $\L\sim$-125\degr, at which point the \citetalias{Antoja2020} sample becomes heavily affected by the contamination of the disc. Remarkably, \citetalias{Antoja2020} find stars in the trailing apocentre amid all the contamination, and the median $\mu_{tot}$ found by these latter authors is compatible with ours (orange shaded area), even though the only criteria used in their selection was based on proper motions. Overall, the differences are of the order of $\sim$0.3 mas\,yr$^{-1}$ at most, and only with more accurate proper motions will we have the precision needed to decipher whether there is a kinematic separation between the different populations within the stream.

We also examined the residuals against LM10, which we show in Panel (e). The agreement is within 2$\sigma$ in the range -150\degr<$\L$<100\degr \ and is only broken near the trailing apocentre, which is expected given the differences between the predicted distances and the observed ones (c.f. Fig.~\ref{fig:LvsDist}), and beyond $L \geq$120$^\circ$, the region where we already know that the Strip selection is poor.

\subsection{Galactocentric distance}\label{sec:GC_dist}
In Fig.~\ref{fig:Sky_vel}, we show a decomposition of the Sgr stream into each of the available dimensions, with the first panel (a) containing the distribution of the candidate stars in the sky for reference. We begin by analysing the trend in Galactocentric distance with $\L$ by converting the heliocentric distances derived in Sect.~\ref{sec:dist} with the parameters of the Sun described in Sect.~\ref{sec:initial_sample}. We are interested in the distribution of the tidal debris across the Galaxy since this gives us a direct insight into the actual orbit of the progenitor (\citealt{BinneyTremaine2008}, but see \citealt{Sanders2013}). Additionally, we also want to estimate the location of the apocentres and their angular separation, as these quantities can be used to characterise the stream and compare with independent determinations and models.

Panel (b) of Fig.\ref{fig:Sky_vel} shows the Galactocentric distance distribution of the stars in the Strip sample as a function of $\L$. In orange we show the smoothed median in bins of 5\degr  and its associated 1$\sigma$ error. The green line corresponds to the prediction by LM10. We notice that the model does not follow our measurements, deferring more than 3$\sigma$ in most parts of the stream, including the centre of the dwarf. These discrepancies cannot be due to the assumed metallicity zero point. We use Eqs.~\ref{eq:LvsZ} and \ref{eq:dist} to convert the residuals in distance observed in the leading arm into differences in metallicity and we find that there should be a change of $\sim$1\,dex in $\sim$50\degr\ to account for it. This jump is much larger than the uncertainties in metallicity and would have been easily recognisable in Fig.~\ref{fig:FeHvsL}. Moreover, the green and orange lines (model and data, respectively) cross at $\L\sim$30\degr \ and again at $\L\sim$80\degr\  implying a change of sign in the residuals. This would also require a shift  in the sign of the offset ($\Delta Z$ defined in Sect.~\ref{sec:dist}) to explain it, which by construction is not possible. Therefore, it is safe to conclude that the deviations from the model are indeed physical. The same applies to the trailing arm, where the discrepancy in metallicity is too large to be artificial. On the other hand, the pattern we find is compatible with the determination made by \cite{Hernitschek2017} although, compared to their work, we do not use any modelling of the halo and, as a result, the trend that we recover is smoother.

We tried to calculate the apocentres using the recipe described in \cite{Belokurov2014b} and used again in \cite{Hernitschek2017}, which consists of describing the apocentres of the debris as a Gaussian profile. We find that fitting a Gaussian to the leading arm does not accurately reproduce our observations and the results are highly dependent on the range of $\L$ used. Instead, we construct a cubic spline (orange line in Fig.\ref{fig:Sky_vel}.b) to smooth the median and obtain a curve that can be evaluated at any point. We then use it to search for the local maximum. Due to the underlying binning, we use half the bin size in $\L$ as the uncertainty in the angular position. In distance, we use the error in the median shown as an orange shaded area in Fig.\ref{fig:Sky_vel}.b. Applying this strategy to the Strip samples leads to the following values (see Appendix~\ref{app:nGC3_summary} for the values estimated with nGC3):

\begin{itemize}
    \item Trailing: $\L=-193^\circ \pm 2\fdg5$ at $D_{GC}=(92.48\pm1.45)$~kpc
    \item Leading: $\L=64^\circ \pm 2\fdg5$ at $D_{GC}=(47.73 \pm0.48)$~kpc,
\end{itemize}

\noindent with an angular distance between apocentres that implies a differential heliocentric orbital precession of $\omega_\odot=360^\circ+\L^T-\L^L=103^\circ\pm3\fdg5$.

We only show the statistical uncertainties but we stress again that an incorrect metallicity zero point would bias these estimates. Nevertheless, we expect the difference to be much smaller than 4\% in distance (Sect.~\ref{sec:dist}). On the other hand, the contribution of the uncertainty in the GC-Sun distance is negligible.

Our determination for the Strip, shown as vertical grey bands across the whole of Fig.~\ref{fig:Sky_vel}, has to be compared with the values derived by \cite{Belokurov2014b}: $\L=71\fdg3 \pm 3\fdg5$ at $D=(47.8 \pm0.5)$~kpc for the leading arm and $\L=-189\fdg5 \pm 1\fdg0$ at $D=(102.5 \pm2.5)$~kpc for the trailing arm ($\omega_\odot=99\fdg3 \pm3\fdg5$). We note that there is a $\sim$3.5$\sigma$ discrepancy at the far apocentre, which cannot be reconciled by tuning $\Delta Z$ without creating tension in the near one. In contrast, the results of \cite{Hernitschek2017} using the RR Lyrae from PS1 are closer to ours: $\L=63\fdg2 \pm 1\fdg2$ at $D=(47.8 \pm0.5)$~kpc for the leading tail and $\L=-192\fdg4 \pm 0\fdg4$ at $D=(98.95 \pm1.4)$~kpc for the trailing tail ($\omega_\odot=104\fdg4 \pm1\fdg3$). Nevertheless, the tension is still present. The major difference is that we have obtained our measurements in a model-independent way and using all the stars in our sample.

\subsection{A peek into the unknown: tangential velocities}\label{sec:tan_vel}
\begin{figure*}
    \centering
    \includegraphics[width=\linewidth]{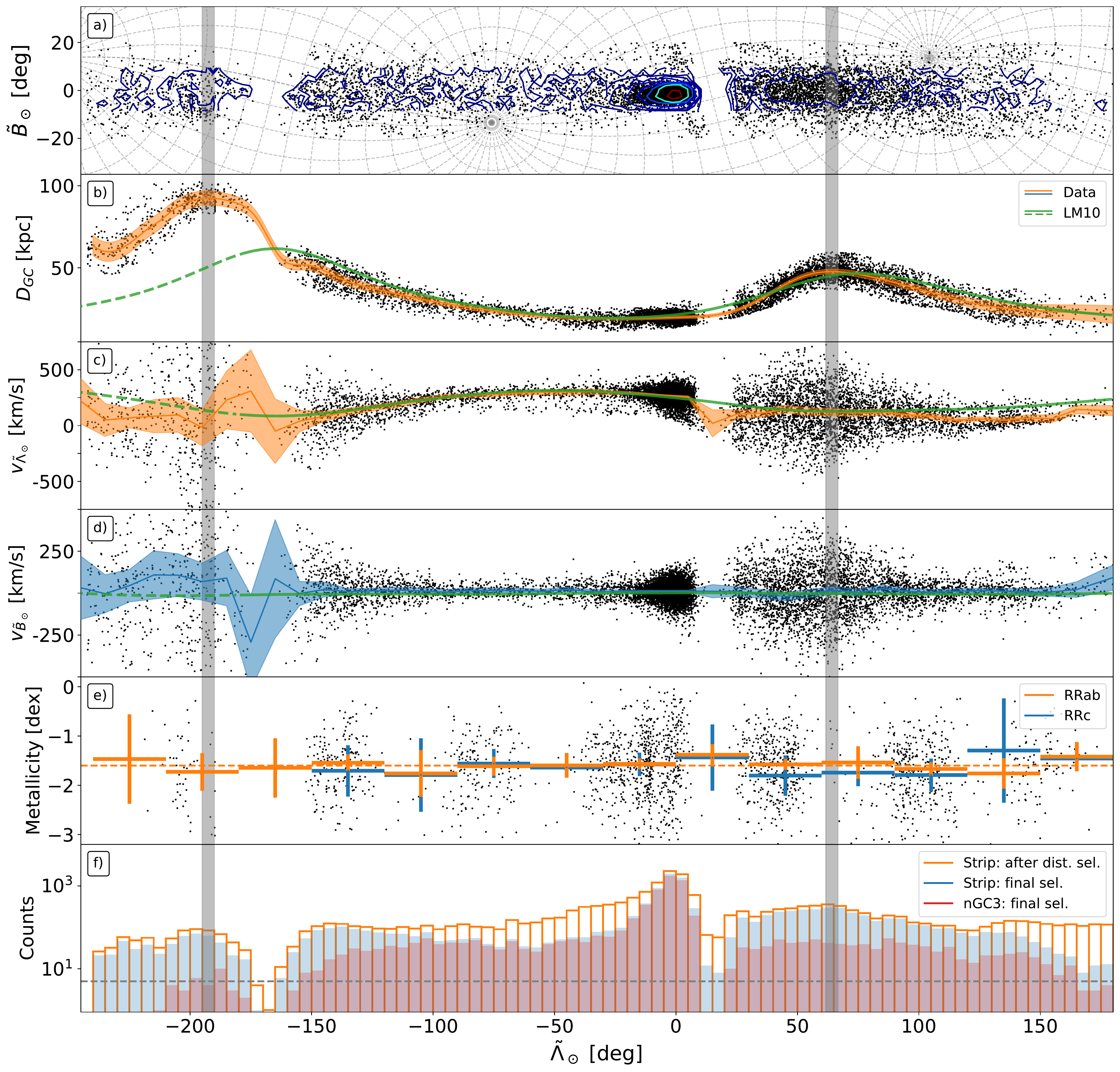}
    \caption{Position, distance, and velocity as a function of $\L$ of the Sgr stream (Strip sample). a) Sky distribution with overlaid count contours. The dashed lines represents a regular grid in the Galactic coordinate frame. b) Smoothed median (orange line) and associated uncertainty (orange area) of the Galactocentric distance for the data (black dots) and LM10 (green line). c-d) Same as (b) but for the velocity in the $\L$ (orange) and $\tilde{B}_\odot$ (blue) directions after correcting for the solar reflex. e) Median metallicity of the RR Lyrae types AB (orange) and C (blue). f) Histograms of $\L$ for the Strip sample -- but only after selecting by distance -- (orange), for the final Strip sample (blue) and for the nGC3 sample (red). The dashed black line represents the threshold below which we do not compute the median used in the other panels. We also show the positions of the apocentres as two vertical grey stripes (see text).}
    \label{fig:Sky_vel}
\end{figure*}

Finally, with all five dimensions together (position, proper motion, and distance) we can access an almost unexplored aspect of the stream: the tangential velocity. To do so, we make use of the \textit{Python} package {\sc GALA} \citep{gala,Price-Whelan2017} which provides a tool to correct for the solar reflex given the position and velocity of the Sun with respect to the GC. For that, we use the parameters described in Sect.~\ref{sec:methods}. The resulting velocities for the Strip sample are shown in panels (c) and (d) of Fig.~\ref{fig:Sky_vel} for the velocity along ($v_{\L}$) and across ($v_{\tilde{B}_\odot}$) the stream, respectively. In contrast with the previous section, here we smooth the median values in bins of 10\degr in $\L$ with a Gaussian filter (similar to \citetalias{Antoja2020}). 

When looking at the velocity along the $\L$ axis (orange line in Fig.~\ref{fig:Sky_vel}.c) we note the general agreement with the LM10 model (green line). In most of the regions, both trends are compatible within 3$\sigma$ except for i) the dwarf itself, where we find that our results are $\sim$20\,km\,s$^{-1}$ above the predictions; and ii) beyond $\L$>\,80\degr, where the differences exceed 70\,km\,s$^{-1}$ at some points. While the latter could be related to the contamination, it would need to bias both the distance by more than 5\,kpc and the proper motions by more than 0.3\,mas\,yr$^{-1}$ to reconcile our observed $v_{\L}$ with the model. A more likely scenario is that the orbit of the progenitor in LM10 is not accurately reproducing the observations, which is consistent with our findings of Sect.~\ref{sec:GC_dist} and with the fact that we observe a significantly different tangential velocity for the dwarf.

The velocity perpendicular to the orbital plane of Sgr is also interesting since it is where the effects of other satellites like the Magellanic Clouds are expected to be more noticeable. This is the case of the Orphan stream, where the misalignment between the velocities of the stars and its orbit have been related to the Large Magellanic Cloud \citep{Erkal2019}. In panel (d) we show the changes in $v_{\tilde{B}_\odot}$ along the stream. We find no significant deviation from zero at the current level of precision but we notice the degeneracy with the velocities used to correct the solar reflex. \citet{Hayes2018} used a sample of Sgr stars to obtained an estimate of the Sun's velocity with respect to the GC by fitting the observations to the model of LM10. However, ideally  we would like to obtain a global fit that can be used to quantify the deviations caused by the satellites and simultaneously estimate the solar parameters. While our sample can indeed be used for that purpose, this analysis is out of the scope of the present work.

We can also obtain a measure of the total velocity projected in the sky  of the Sgr dwarf that can be used to initialise simulations from the present time, for example. We do so by summing in quadrature both components of velocity for all stars within a cone of 5$^\circ$ around the centre of the Globular Cluster M54 \citep{Bellazzini2008}. Then, the median tangential velocity is $V_\perp = 272.56\pm0.02$\,\kms{}.

Our final output of this part is an interpolator for each component of the velocity for any $\L$ along the stream. Despite the large uncertainties in proper motions, we can still recover a trend that can be used to test future models of Sgr. In Appendix \ref{app:interpolators} we explain how to download them, along with the rest of the interpolators used in this work, such as distance, proper motion, and so on.

\subsection{Metallicity}\label{sec:metallicity}
Now that we have a reliable sample of candidates, we can check our initial assumption of a constant metallicity along the arms of Sgr (Sect.~\ref{sec:dist}). The fifth panel of Fig.~ \ref{fig:Sky_vel} shows the metallicity distribution for the stars for which this parameter has been measured. The error bars correspond to the median, with the associated error, in bins of 30$^\circ$ for both RRab (orange) and RRc (blue)  RR Lyrae. The first thing we note is that the median oscillates around the dashed line that represents the mean value calculated in Sect.~\ref{sec:dist}, with all stars in the range of distance between 20\,kpc and 50\,kpc: -1.6 dex. 

With this we can already clearly see that there is no appreciable gradient. Nevertheless, we measure it by fitting a line to all RRab stars in the leading ($\L$>0\degr) and the trailing ($\L$<0\degr) arms. By doing so we obtain a slope of (-1.8$\pm$0.5)\,$\times 10^{-3}$\,dex\,deg$^{-1}$ and (0.3$\pm$0.4$)\,\times 10^{-3}$\,dex\,deg$^{-1}$ for the leading and trailing arms, respectively. 
While this could be easily disregarded as unimportant, we note that the current estimates of the metallicity gradient in the tails of Sgr are of the order of 10$^{-3}$\,dex\,deg$^{-1}$ \citep{Hayes2019}. Nevertheless, considering the inhomogeneous coverage of the sky of the subsample of stars with metallicity and the possible contribution from contamination, it is hard to trust the variations that we measure. In the worst case scenario, this would cause a difference at the trailing apocentre of $\sim$0.2\,dex, which is of the order of the precision achievable with photometric metallicities. In turn, such a difference would imply that our measurement of the distance obtained in the previous section could be short by $\sim$3\%, that is, that the trailing apocentre could alternatively be at a distance of $\sim$96\,kpc instead of $\sim$93\,kpc. While this effect would bring our estimate closer to the values in the literature\footnote{If we consider the gradient in both tails, then the estimate of the distance at the leading apocentre grows to $\sim$48.5\,kpc, departing from the values estimated by previous authors.}, it is still within its 2$\sigma$ confidence interval. In summary, we can say that there is no measurable metallicity gradient at the current level of precision offered by the photometric metallicities.

Finally, we show the mean values of metallicity for each type of RR Lyrae star in Table~\ref{tab:metallicity} from which we can conclude that the metallicity of this sample is -1.61$\pm$0.01\,dex.

\begin{table}
\caption{Metallicity statistics of the nGC3 (Strip) final sample. For each type of RR Lyrae (first column) we show: the number of sources classified as such with reported metallicity (second column), their mean metallicity as given in the \Gaia archive (third column), and the associated standard deviation (fourth column).}\label{tab:metallicity}
\centering
\begin{tabular}{cccc}
\hline\hline
Type & N & <[Fe/H]> [dex] & $\sigma_{[Fe/H]}$ [dex]\\
\hline
RRab & 997 (1\,876) & -1.60 (-1.62) & 0.63 (0.62)\\
RRc & 89 (149) & -1.62 (-1.65) & 0.45 (0.46)\\
RRd & 1 (2) & -1.11 (-1.10) & - (0.02)\\
\hline
\hline
\end{tabular}
\end{table}

\subsection{Completeness assessment}\label{sec:counts}
Here, we inspect the distribution of the counts along the stream and the overlap between the different samples.
The bottom panel of Fig.~\ref{fig:Sky_vel} shows the number of stars of the final Strip sample (blue bars) in 5$^\circ$ bins compared to the counts after the selection based on distance of Sect.~\ref{sec:members} (orange step). Here, the dashed black line represents the imposed limit of five stars below which we do not compute the median used in panels (b), (c), and (d). 

As expected, a significant fraction of the stars fall inside the dwarf itself: 4\,730 (40\%) in the region $\L \in$ [-10$^\circ$,\,10$^\circ$]. The rest are spread rather uniformly except for the two parts where the stream crosses behind the MW disc, which are easily identifiable by a sudden drop in the counts. We do observe a larger number of stars in the leading arm compared to the symmetric $\L$ in the trailing arm, which is to be expected near apocentre where the stars in the stream slow down. Nevertheless, as explained in Sect.~\ref{sec:pmra}, the increase of the proper motion uncertainties near apocentre prevents us from effectively reducing the contamination there. That is why the orange (Strip sample after selecting only based on distance) and blue (final Strip sample) histograms are so close to each other in that region.

We also include the counts of the final nGC3 sample as red bars. In this case, the two distributions are similar for $\L$<0$^\circ$ but are more different for the leading arm. The cross-match between both final samples tells us that all but 44 stars in the nGC3 sample are in the Strip sample. Thus, the missing stars in the leading tail are those that cause the difference in total proper motion seen in Fig.~\ref{fig:ProperMotion}. Comparing both samples, we note that \cite{Cseresnjes2001} estimated the total amount of RRab stars to be $\sim$8\,400. This value is almost exactly halfway between the number of stars in the nGC3 sample and the number in the Strip sample, reinforcing the idea that the former is pure but incomplete and the latter is complete but contaminated.

Regarding the cross-matched with \citetalias{Antoja2020}, the first thing that is important to mention is that the sample of \citetalias{Antoja2020} contains only 3\,476 RR Lyrae of the 182\,495 from the list presented in Sect.~\ref{sec:data}. Compared to the Strip sample, the sample used in \citetalias{Antoja2020} contains 3\,233 of the 11\,677 selected in this work. If we restrict ourselves to the region where the contamination is lower (-120$^\circ$<$\L$<150$^\circ$) then we find that out of the 3\,145 RR Lyrae in \citetalias{Antoja2020}, 3\,033 are among the 10\,431 of the Strip sample. This means that the recovery fraction is around 95\%. If we now analyse the cross-match between the nGC3 and \citetalias{Antoja2020} samples, we see that the sample of \citetalias{Antoja2020} contains 2\,712 stars of the 6\,608 in the nGC3 sample; 2\,661 out of 6\,386 stars if we only consider the low contamination region defined in Sect.~\ref{sec:counts}. The resulting recovery fraction is around 80\% for the nGC3 sample. Since we obtained these estimates from low-contamination regions, we expect them to be similar to the true completeness but we stress that these percentages apply only to the true members among all the candidates (as if there was no contamination). In any case, the RR Lyrae in \citetalias{Antoja2020} that are \emph{not} in any of our two samples can mostly be found in the dwarf, the tip of the tails (where we know that the contamination in \citetalias{Antoja2020} is higher), and, most interestingly, the leading apocentre. However, further investigation of the differences among the samples requires a fine characterisation of the different selection functions and we leave this for a future study.
On the other hand, the recovery fraction for the RR Lyrae in \citetalias{Antoja2020} is roughly between 30\% and 40\% (we cannot extrapolate this estimates of completeness to stars with other colours and magnitudes). 
Nevertheless, expanding the list of \citetalias{Antoja2020}  would require a good treatment of the contamination and probably the introduction of a model for the halo and thin/thick disc.  

\subsection{Velocity dispersion}\label{sec:vel_disp}

As mentioned in \citet{Gibbons2014}, the stars that escape through the Lagrange points do so with a dispersion in velocity inherited from the dispersion that the stars had at the outskirts of the dwarf, the latter being proportional to the total mass of the infalling satellite. Additionally, the stream can become kinematically heated by the continuous interaction with dark matter subhalos \citep{Johnston2002}. Therefore, learning about the velocity dispersion of the stream allows us to obtain a tighter constraint on the dynamical mass of the progenitor and potentially a test for dark matter models.

\begin{figure}
    \centering
    \includegraphics[width=\linewidth]{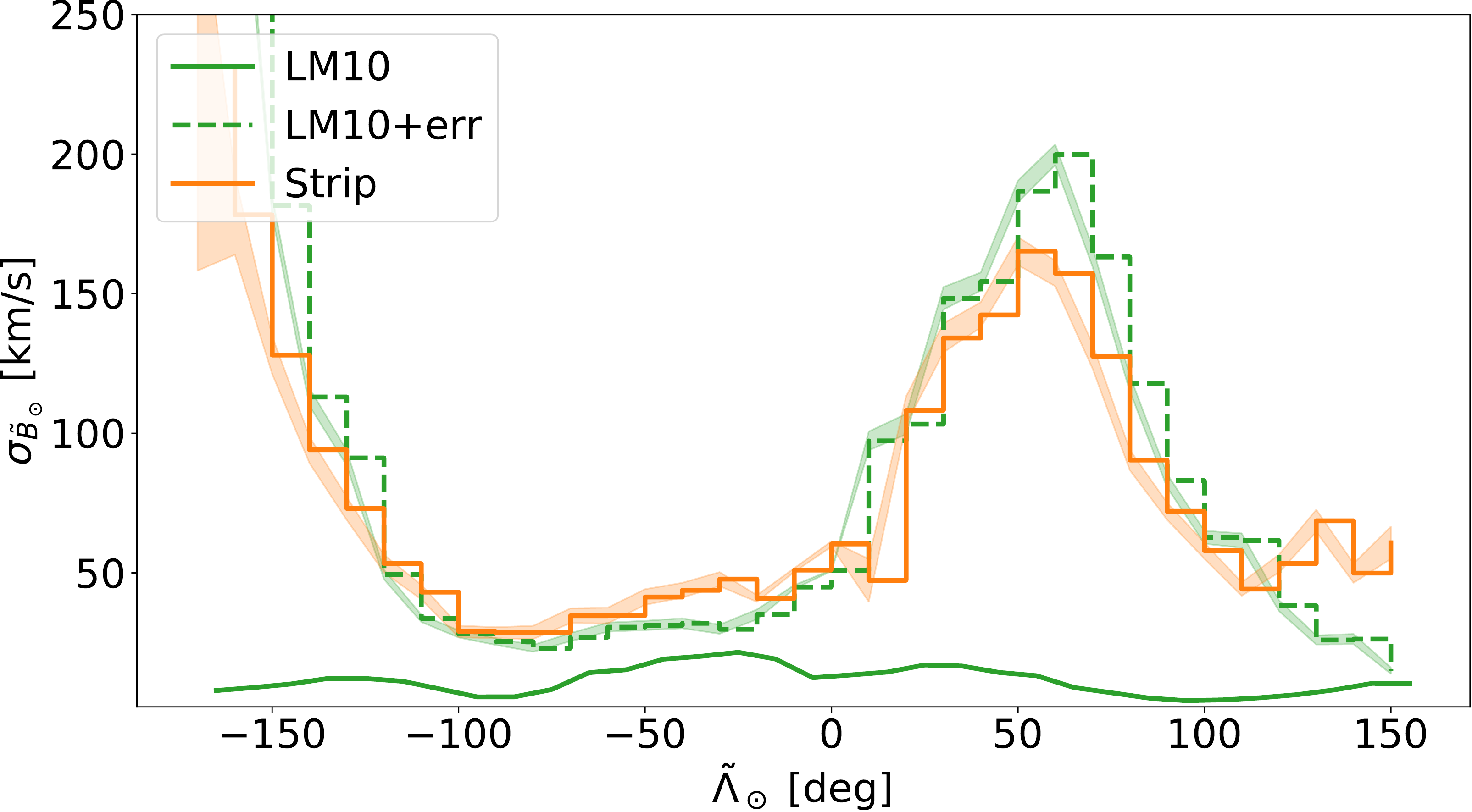}
    \caption{Velocity dispersion profile for $v_{\tilde{B}_\odot}$ as a function of $\L$. The orange steps correspond to the velocity dispersion observed in the final Strip sample. The green lines are the LM10 model: the solid is for the model particles and the dashed for the model after introducing the observed errors in proper motion.}
    \label{fig:vel_disp}
\end{figure}

In Fig.~\ref{fig:vel_disp} (Fig.~\ref{fig:vel_disp_ngc3}) we show the velocity dispersion profile of $v_{\tilde{B}_\odot}$ for the Strip (nGC3) sample. We also include the profile of the LM10 model in green (solid line). We have been able to reproduce both curves easily with the LM10 particles by randomly assigning the uncertainties of the observed stars and subsequently applying only the initial selection (Sect.~\ref{sec:initial_sample}). The remaining differences between the observed profiles and the ones predicted by the model are mostly due to the fact that we did not apply the whole methodology described in Sects.~\ref{sec:members} to \ref{sec:pmdec}. If we wanted to obtain an estimate of the deviations caused by possible subhalos, it would be necessary to build the selection function of our two samples and accurately model the observational errors.

\subsection{The bifurcation}\label{sec:bifu}
\begin{figure}
    \centering
    \includegraphics[width=\linewidth]{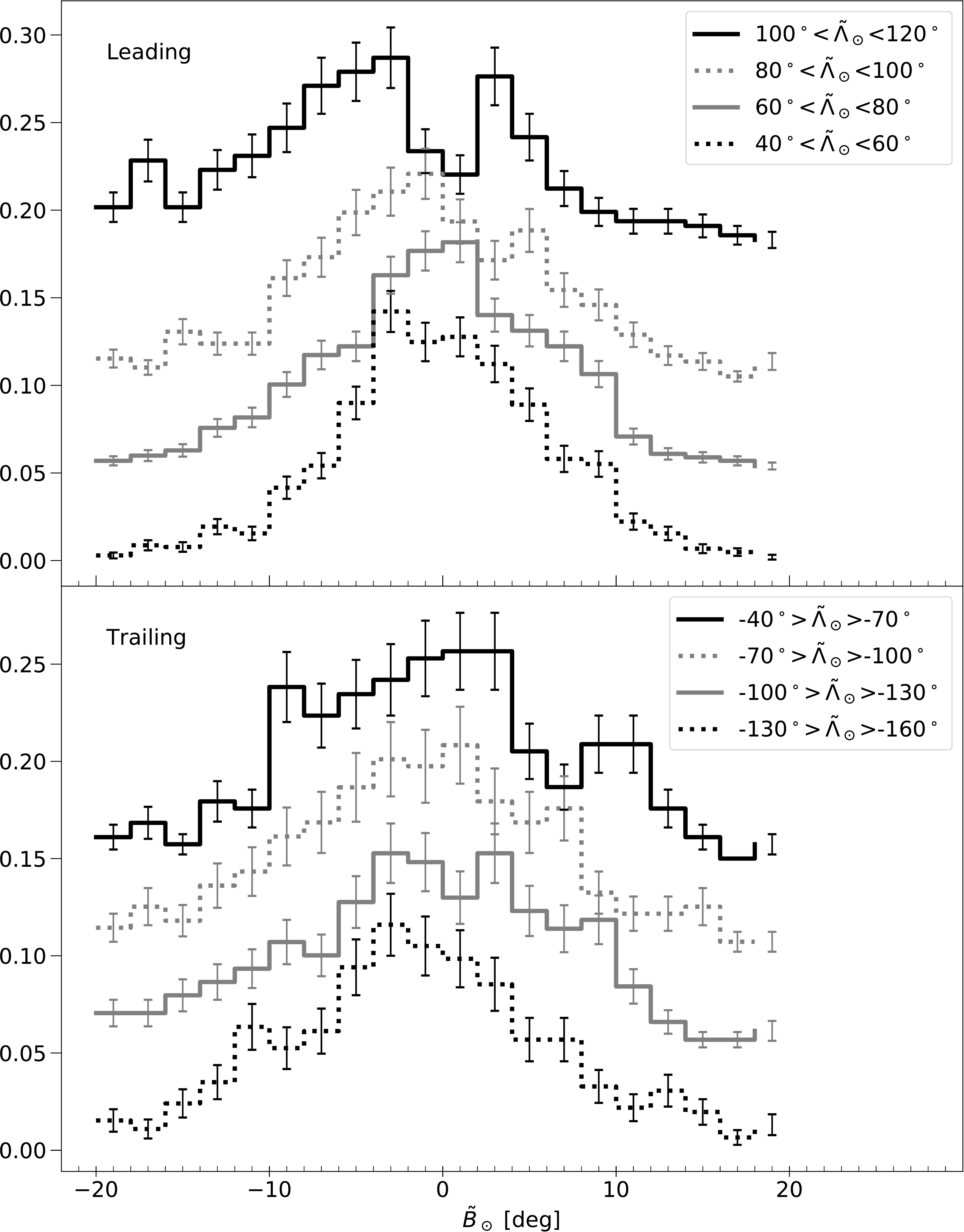}
    \caption{Histograms of $\tilde{B}_\odot$ normalised to the number count for different bins of $\L$ for the Strip final sample. Top: Bins in the leading arm. Bottom: Bins in the trailing arm. We have added an arbitrary offset in the y-axis to each curve for visualisation purposes.}
    \label{fig:bifu_strip}
\end{figure}
The bifurcation of the Sgr stream was observed for the first time by \cite{Belokurov2006} in the leading tail, and later detected in the trailing arm by \cite{Koposov2012} and \citet{Navarrete2017}. This feature can be clearly seen with the M giants and shows the stream splitting into two branches creating a bi-modal distribution of $\tilde{B}_\odot$ in some portions of the tails. Although some models have been proposed to explain this feature with only one progenitor, for example \citet{Penarrubia2010} (but see \citealt{Penarrubia2011}), there is no consensus on its cause. Thus, it is important to provide additional observational evidence with different tracers to constrain the range of possibilities.

Figure \ref{fig:bifu_strip} shows the histogram of $\tilde{B}_\odot$ of the final Strip sample for different bins in $\L$ within the range -160$^\circ$<\,$\L$<120$^\circ$ (again, see Appendix~\ref{app:nGC3_summary} for the equivalent plot for the nGC3 sample). In some portions of the leading arm (top panel) a bi-modal distribution is evident. The strongest signal is in the range 100$^\circ$<$\L$<120$^\circ$, where the two highest peaks are separated by $\sim$6$^\circ$. We note the agreement between our findings and those in \cite{Belokurov2006}, where these latter authors report the beginning of the bifurcation at $\L\sim$80$^\circ$ and the separation between the two branches growing from $\sim$7$^\circ$ to $\sim$10$^\circ$ at $\L\sim$120$^\circ$. In the trailing arm, there is no  appreciable bifurcation. We have increased the bin width to compensate for the lower number of counts, but still we cannot see a significant signal (bottom panel of Fig.~\ref{fig:bifu_strip}). 

\begin{figure*}
    \centering
    \includegraphics[width=\linewidth]{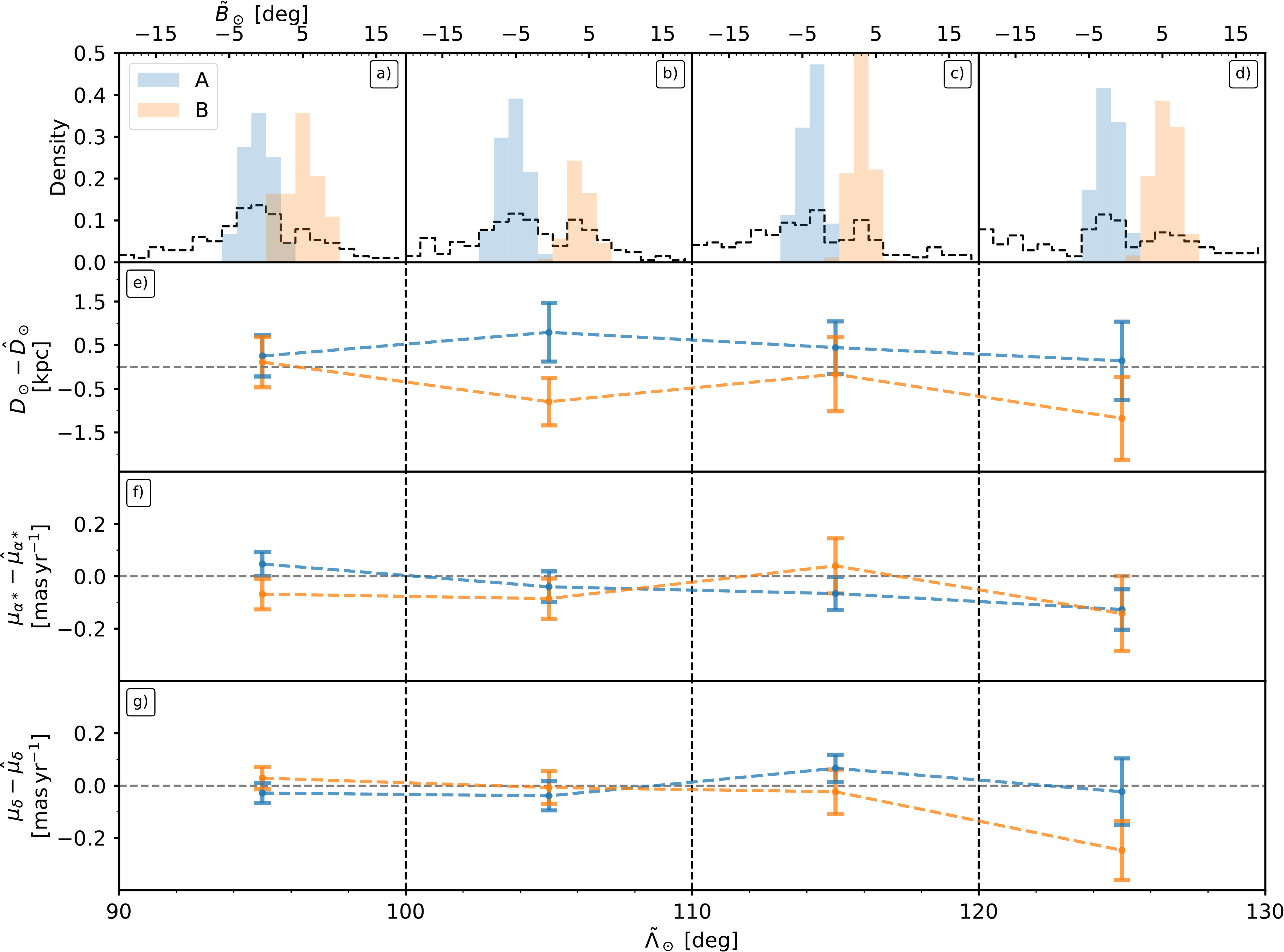}
    \caption{Comparison between the bright (A, blue) and faint (B, orange) branches of the bifurcation at different $\L$>90$^\circ$ (Strip sample). The top panels (a) to (d) show the histogram of $\tilde{B}_\odot$ (black line) corresponding to the stars that fall in the range of $\L$, from left to right, [90$^\circ$-100$^\circ$, 100$^\circ$-110$^\circ$, 110$^\circ$-120$^\circ$, 120$^\circ$-130$^\circ$]. We also show the weighted histograms for each branch, which highlight  the stars that contribute to compute the means plotted in the panels below. 
    In panel (e) we plot the weighted mean heliocentric distance, after subtracting the median distance, and the associated error. 
    Panels (f) and (g) contain the weighted mean and its error for the proper motion in right ascension and in declination, respectively. Here we also subtract the median value of the bin to cancel the overall gradients.
    }
    \label{fig:bifu_detail}
\end{figure*}

To investigate the possible differences between the two branches observed in the leading arm, we compute the weighted means for the distance and proper motions in four bins of $\L$: 90$^\circ$ to 100$^\circ$, 100$^\circ$ to 110$^\circ$, 110$^\circ$ to 120$^\circ$, and 120$^\circ$ to 130$^\circ$. The weight assigned to each star represents the probability of belonging to one branch or the other based on their $\tilde{B}_\odot$. To obtain these two probability distribution functions, first we create a kernel of $\tilde{B}_\odot$ to have a continuous representation of the histogram. We then locate the only two local maxima that the kernel has and associate each one with either branch A (negative $\tilde{B}_\odot$) or branch B (positive $\tilde{B}_\odot$). After that, we restrict ourselves to $\pm$1.5\degr of one of the peaks, that is,,  we only use the values of the kernel near the local maxima, and fit a parabola with which we can obtain a probability at every $\tilde{B}_\odot$ (wherever the parabola returns a negative probability we instead use zero). Finally, we repeat the procedure for the other peak to obtain the second probability distribution. Panels (a) to (d) of Figure \ref{fig:bifu_detail} show the original histogram of $\tilde{B}_\odot$ at each bin in $\L$ (black line) along with the resulting weighted histograms for branches A (blue) and B (orange). Panels (e) to (g) contain the corresponding weighted averages and their 1$\sigma$ uncertainties in heliocentric distance, proper motion in right ascension, and proper motion in declination, respectively. To compute the uncertainty in this estimate, that is, the standard error of the weighted mean (SEM), we used the formalism of \citet{Cochran1977}\footnote{$(SEM_w)^2 = \frac{n}{(n-1)\sum \omega_i}[\sum(\omega_i x_i - \bar{\omega}\bar{x}_w)^2-2\bar{x}_w\sum(\omega_i-\bar{\omega})(\omega_i x_i - \bar{\omega}\bar{x}_w) + \bar{x}_w^2\sum(\omega_i-\bar{\omega})^2]$}. In order to cancel the gradients observed in the previous sections, we subtract the median value obtained from the stars within the bin.

The result is that the differences that we measure between the two sides of the bifurcation are always smaller than 2$\sigma$. Nevertheless, we note that the orange line in panel (e) is always below the blue one. This can be due to many factors, such as projection effects caused by the stream being a 3D structure, or contamination, since the faint branch (B) is more susceptible than the bright branch (A), or simply because one branch is actually more extincted than the other, which may not be well reproduced with the extinction map that we use. As an example, supposing that the latter is true, we would only need to add <\,0.1 mag of extinction to the mean distance of branch A to bring the blue line in panel (e) down to the orange line. 

In contrast with the recent study of \citet{Yang2019}, where these latter authors report differences in velocity and metallicity for Blue Horizontal Branch stars in the bifurcation, we conclude that there is no significant separation in either kinematics or distance between the two branches of the leading tail. Furthermore, if the branches were indeed different, their separation should be of the order of <1.5\,kpc and <0.2\,mas\,$^{-1}$. We have also looked at the metallicity and find no significant difference, but due to the small number of stars available and the measurement errors, we cannot give any meaningful upper limit. Finally, the fact that we do not observe the bifurcation in the trailing tail could be related either to our selection function or simply to a lack of stars.

\section{Conclusions}\label{sec:conclusions}
\Gaia has revolutionised the field of Galactic astronomy. With only 22 months of data, the scientific community has already been able to make numerous and quite significant discoveries. Furthermore, DR2 can produce novel results concerning areas that have already been thoroughly examined. The Sgr stream is a good example of this, and in this work, as well as in \citetalias{Antoja2020}, we wanted to demonstrate what its discovery would have been like in the \Gaia era, using only its kinematics.

We have produced two catalogues of probable RR Lyrae members of the Sgr dwarf and stream, starting with the most basic assumptions and exhausting the \Gaia outputs to carefully select the list of candidates step by step. The two catalogues are produced by running this trimming process twice, each time from a different starting point. In the first case, we look for stars that are rotating coherently in a plane through the GC and end up with the nGC3 sample: a higher-purity but lower-completeness ($\approx$\,80\%) sample. By  demanding that the stars and their velocities do not deviate significantly from the mean orbital plane, we are able to efficiently discard most of the contamination at the cost of introducing kinematical biases. The second time, we select all stars in the sky that are within $\pm$20\degr \ from the  known mid-plane of the stream. In doing so, we obtain the Strip sample: a higher-completeness ($\approx$\,95\%) but lower-purity sample. In this case, we rely on our data-driven selection schema based on distances and proper motions to get rid of most of the contamination. We have composed a list of 11\,721 stars belonging to either the Strip (11\,677 stars) or the nGC3 (6\,608 stars) sample, and have made this list publicly available. 

The main output of this work is the largest sample of RR Lyrae in the Sgr stream to date with both proper motions and distances together. This allows us to study, for the first time, the tangential velocities along the whole stream and provide a reference to compare against.  Our results confirm the findings by \citetalias{Antoja2020} regarding the proper motion track along the stream. We observe more or less the same trends in proper motion, with differences of less than 0.3 mas/yr in most parts of the stream, with the only exception being  those regions where either our sample or that of \citetalias{Antoja2020} are known to be contaminated. Also, we note that 93\% of the RR Lyrae contained in \citetalias{Antoja2020} are inside our final sample, confirming that both are detecting the same structure. It will be interesting to compare our work with that of \cite{Ibata2020} in which, as we recently found out, they also provide a list of RR Lyrae obtained with \textit{STREAMFINDER} \citep{Malhan2018}. In doing so, we can produce an independent estimate of the completeness and quantify the possible biases. Of particular interest is the \emph{Sagittarius Stream Selected Sample} of \cite{Bellazzini2020}, composed of the 5\,385 RR Lyrae from the \begin{tt}gaiadr2.vari\_rrlyrae\end{tt} catalogue that pass a set of filters requiring that the stars are near the orbital plane of Sgr, rotate with it, and have a $\mu_{\L}$ trend compatible with the fit obtained by \cite{Ibata2020}. This sample is, by construction, very similar to our nGC3 since it requires that the stars must have their velocity vectors aligned with $\L$. Nevertheless, apart from having less stars, we note from  Fig.~1 of this latter-mentioned publication that there is no specific cut in distance. 
This means that on the one hand the nearby stars close to the bulge have not been removed, while on the other, features like the 
\emph{fluff} \citep{Sesar2017c} can be better preserved.

We compare the evolution of the different observables with $\L$ against LM10, the reference model for most studies of Sgr. The residuals in proper motion are within the uncertainties, apart from the known issues discussed in Sect.~\ref{sec:properties}. That is, that the Strip's mean $\mu_{\alpha*}$ is not reliable beyond $\L$>120\degr and that the nGC3 sample is biased. When comparing the heliocentric distances, we find that the discrepancies are very much significant and not attributable to a gradient in metallicity or an incorrect calibration. Instead, we confirm the results of previous works (e.g. \citealt{Belokurov2014b,Hernitschek2017}) which already pointed out that the apocentres in LM10 are inaccurate in terms of both $\L$ and distance. More recent models are able to reproduce this, such as the one by \cite{Fardal2019},  but  how well their proper motions perform in light of the new observational constraints that we provide here is yet to be tested. We have also determined new apocentres based on the maximum Galactocentric distance of the tidal debris: $\L=-193^\circ \pm 2\fdg5$ at $D_{GC}=(92.48\pm1.45)$~kpc for the trailing tail and $\L=64^\circ \pm 2\fdg5$ at $D_{GC}=(47.73 \pm0.48)$~kpc for the leading. Our values are in some tension with those determined by \cite{Hernitschek2017} which cannot be explained by an incorrect metallicity zero point, and are most likely related to the fact that we used a completely empirical method to make our estimates. 

With the Strip sample, we have been able to test the tangential velocities predicted by LM10. We find that the predictions agree within uncertainties in the trailing arm but fail to reproduce our observations at the location of the dwarf and at the leading arm. Moreover, we provide a measurement of the total velocity projected in the sky for the progenitor, $V_\perp$ = (274.18$\pm$0.02) \kms{}. Despite the complexity of our selection process, we can add that the velocity dispersion profiles of the model are also compatible with our observations. 

Regarding the metallicity, we do not observe any meaningful gradient with $\L$ at the current level of precision and the mean value we recover is -1.6 dex in the ZW84 scale. This is in contrast with other studies where  differences are detected for most populations. In particular, \cite{Yang2019}  reported a significant difference between the leading and the trailing arm for the Blue Horizontal Branch stars. Even though we use photometric metallicites instead of spectroscopic ones, the former having larger uncertainties, and the hypothetical gradient has to be of the order of 10$^{-3}$\,dex\,deg$^{-1}$ at most in absolute value. Such small gradients would cause differences along the stream no larger than $\sim$0.2\,dex, which is similar to the typical precision for the photometric metallicities. However, we notice that the gradients that we have measured, which are compatible with the values obtained by \cite{Hayes2019}, would imply that the trailing arm is on average slightly more metal rich than the leading arm, in agreement with previous works (e.g. \citealt{Yang2019}).

The nGC3 sample is a great starting point for a spectroscopic follow-up. Taking advantage of its high purity, the risk of measuring abundances and radial velocities for an off-stream star is minimised. The resulting catalogue would be the most extensive 6D+abundance sample of Sgr RR Lyrae to date, providing a more detailed picture of the metallicity of the stream and probing the full phase-space information for a single tracer.

We also searched for the bifurcation in our data and indeed observe it in both samples. We see a clear signal in the leading arm where it was originally discovered. We observe no significant separation in velocity, distance, or metallicity between the bright and faint branches. Also, we do not detect the bifurcation on the trailing arm where it was found by \citet{Koposov2012}, perhaps because of the low number of stars we have in that region or because of the filters in distance and proper motion used to select the members. It will be interesting to obtain follow-up spectroscopic observations of the stars at each branch and study whether  or not there is a separation in line of sight velocities as well.

Finally, we provide empiric trends (smoothed medians) in distance, proper motion, and tangential velocities of our final Strip sample (see Sect.~\ref{app:interpolators}), along with the respective uncertainties. We also give these trends as interpolators that can be evaluated at any $\L$ , which can be very helpful when modelling the stream.

\begin{acknowledgements}
We thank the anonymous referee for the suggestions and remarks. We thank also Bertrand Lemasle for the useful comments and discussions, as well as the whole \textit{Gaia} team at Barcelona and Groningen. This work has made use of data from the European Space Agency (ESA) mission {\it Gaia} (\url{https://www.cosmos.esa.int/gaia}), processed by the {\it Gaia} Data Processing and Analysis Consortium (DPAC,\url{https://www.cosmos.esa.int/web/gaia/dpac/consortium}). Funding for the DPAC has been provided by national institutions, in particular the institutions participating in the {\it Gaia} Multilateral Agreement. This project has received funding from the University of Barcelona's official doctoral program for the development of a R+D+i project under the APIF grant and from the European Union's Horizon 2020 research and innovation programme under the Marie Sk{\l}odowska-Curie grant agreement No. 745617.  This work was supported by the  MINECO (Spanish Ministry of Economy) through grants ESP2016-80079-C2-1-R (MINECO/FEDER, UE) and ESP2014-55996-C2-1-R (MINECO/FEDER, UE) and MDM-2014-0369 of ICCUB (Unidad de Excelencia 'María de Maeztu'). This project has received support from the DGAPA/UNAM PAPIIT program grant IG100319. The work reported on in this publication has been fully or partially supported by COST Action CA18104: MW-Gaia. CM acknowledges support from the ICC University of Barcelona visiting academic grants and thanks the \Gaiaf~UB team for hosting her during part of this research. CM also thanks PDU en Ciencias F\'isicas at CURE (Rocha), for their hospitality.

\end{acknowledgements}

\bibliographystyle{aa}
\bibliography{Biblio}

\begin{appendix}
\section{Reddening to absorption transformations}\label{app:reddening}

\begin{figure}
    \centering
    \includegraphics[width=\linewidth]{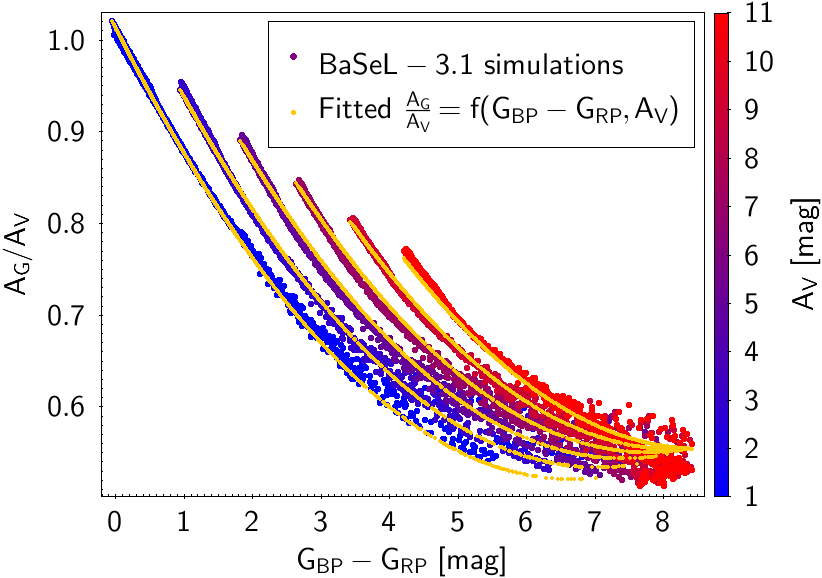}
    \caption{Fitted relationship between $\frac{A_G}{A_V}$ and the observed $G_{BP}-G_{RP}$ colour (affected by reddening) derived from simulations using BaSeL-3.1 synthetic spectral energy distributions \citep{basel31} and \Gaiaf~DR2 passbands from \citet{Evans2018}.}
    \label{fig:AGAVvsBPRP}
\end{figure}

The reddening maps provided by \cite{Schlafly2011} return a value of the colour excess, E(B-V), for each pair of celestial coordinates. To obtain the absorption we then assume a proportionality constant, $R_V$, of 3.1 \citep{Cardelli1989}. However, this yields the absorption in the V-band, $A_V$, and so we still need a conversion to the absorption in the G-band, $A_G$. For that, we use a fitted relationship between the ratio $\frac{A_G}{A_V}$ and the \textit{Gaia} photometry (see Fig.~\ref{fig:AGAVvsBPRP}).

The $\frac{A_G}{A_V}$ ratio is fitted against the $G_{\rm BP}-G_{\rm RP}$ colour using simulations derived from BaSeL-3.1 \citep{basel31} spectral energy distributions using the same methodology as in \cite{Jordi2010}. 

We used the \Gaia passbands derived by \cite{Evans2018} to convert the output of said simulations into observed colours ($G_{\rm BP}-G_{\rm RP}$). Similarly, for the Johnson $V$ passbands we used the response curve provided by \cite{bessell2012}. 

To simulate reddened sources in the range $0<A_{550}=11$~mag, we use \cite{Cardelli1989} extinction law and $R_V=3.1$,  $A_{550}$ being the magnitude of the extinction at a wavelength equal to 550\,nm.

To achieve a better fitting on $\frac{A_G}{A_V}$ we consider a second-degree dependency with the reddened $G_{\rm BP}-G_{\rm RP}$ colour (the observed one) plus two terms depending on the interstellar absorption in the $V$ passband ($A_V$) as well as a crossed term. The obtained relationship is the following:

\begin{eqnarray}
    \frac{A_G}{A_V}&=&0.97883-0.14365\ (G_{\rm BP}-G_{\rm RP})_{\rm obs}\nonumber\\
    &+&0.011077\ (G_{\rm BP}-G_{\rm RP})_{\rm obs}^2+0.034842\ A_V\nonumber\\
    &-&0.0041448\ A_V\cdot(G_{\rm BP}-G_{\rm RP})_{\rm obs.}
\end{eqnarray}

\section{Tailored Gaussian mixture}\label{app:TGM}
Here we explain the TGM algorithm in more detail. We use the particular case of the distance (see Sect.~\ref{sec:members}) to guide the exposition, but the steps are the same for any other quantity that varies as a function of $\L$ ($\mu_{\alpha*}$ and $\mu_{\delta}$). We only tweak the parameters to obtain a better performance for each case: bin size, step size, threshold, maximum separation and kernel bin width.

\begin{figure}
    \centering
    \includegraphics[width=\linewidth]{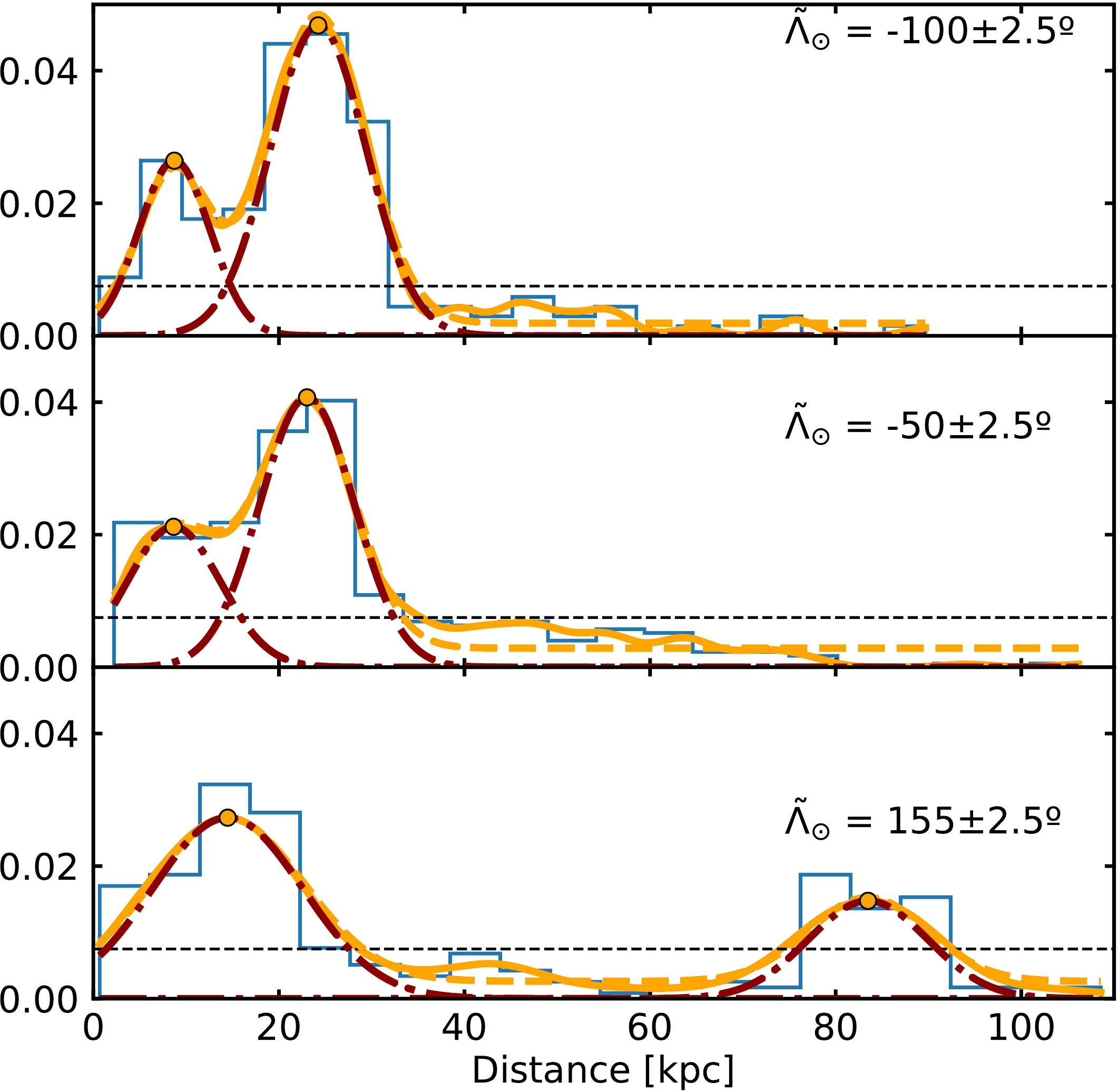}
    \caption{Example of the TGM applied to the distance distribution at selected bins of $\L$=[-100$^\circ$,-50$^\circ$,155$^\circ$] for the Strip sample. The blue steps show a histogram of the heliocentric distance of all stars with a bin size of 5$^{\circ}$. The solid orange line is the kernel obtained with the parameters detailed in the text. The orange dots that appear on top of it are the peaks detected and the red lines the corresponding Gaussian distributions obtained from the fit. Finally, the orange dashed line is the reconstruction of the kernel with the individual Gaussians added together.}
    \label{fig:TGM_example}
\end{figure}

First we take all the stars in the range -2.5$^{\circ}$ $\leq\L<$ 2.5$^{\circ}$ and obtain a Gaussian kernel to the corresponding distance histogram. Having done that, we evaluate the kernel at 100 000 points and apply a peak finder algorithm (\begin{tt}scipy.signal.find\_peaks\end{tt}) to locate all local maximum. We keep only those whose height are above a certain dimensionless threshold, set arbitrarily so that we can remove the small oscillations appearing on the tails of the distributions due to Poisson noise (0.0075 in this case). The remaining $n$ peaks are then used as centres of $n$ normal probability distributions. Finally, we fit the sum of a constant floor level (free parameter) and $n$ Gaussians for which we set free their widths ($\sigma$ [kpc]) and amplitudes. For that, we use the Least Square Minimisation algorithm implemented in \begin{tt}Scipy\end{tt} and find the $\sigma$ for each component that better reproduce the kernel. In order to avoid spending too much time at each bin, we limit the number of Gaussians to the four with highest height. Three examples of the fitting process are shown in Fig.~\ref{fig:TGM_example}.

Given that at $\L$=0$^{\circ}$ the dominant component is that of the Sgr dwarf, we assign the Gaussian with the largest amplitude to the stream. Subsequently, we simply repeat this strategy every 5$^\circ$ in both directions, negative and positive $\L$, but now associating to the stream the component most similar to the previous one. To define which is the most similar we apply the following rule: rank the peaks by height, filter out those that are too far apart, and then chose the first one. In this case, we impose a maximum difference between two consecutive peaks of 10\,kpc. 

This method relies on the separation between the different components at any given $\L$. If they are too close to each other, we need a smaller bin width when creating the kernel to resolve them at the expense of increasing the number of peaks created by simple Poisson noise. On the other hand, if we smooth the kernel too much in an effort to reduce the spurious peaks, the different components can interfere with each other, and therefore we obtain a biased position. Here, we have chosen a value of 0.15 (in the units given by the \begin{tt}Scipy\end{tt} function) for both the Strip and the nGC3 samples.

In Sect.~\ref{sec:pmra} and \ref{sec:pmdec}, we applied the same procedure but with the following parameters:
\begin{itemize}
    \item Bin size: 2.5\degr
    \item Step size: 5\degr
    \item Bin width: 0.30
    \item Threshold: 0.10
    \item Separation: 1.5 mas\,yr$^{-1}$
\end{itemize}

As a final remark, it is also worth mentioning that we tried using the common clustering algorithms like DBSCAN, hierarchical DBSCAN or $k$-means. Nevertheless, most of this methods are based on density  and, in our case, the Sgr stream presents significant gradients which result in suboptimal performance. We acknowledge that there could exist a combination of parameters and a coordinate transformation such that a clean separation between the contamination and the stream stars appears but, after some trial and error, we decided to apply the TGM. 

\section{Strip sample selection process}\label{app:strip}
\begin{figure}
    \centering
    \includegraphics[width=\linewidth]{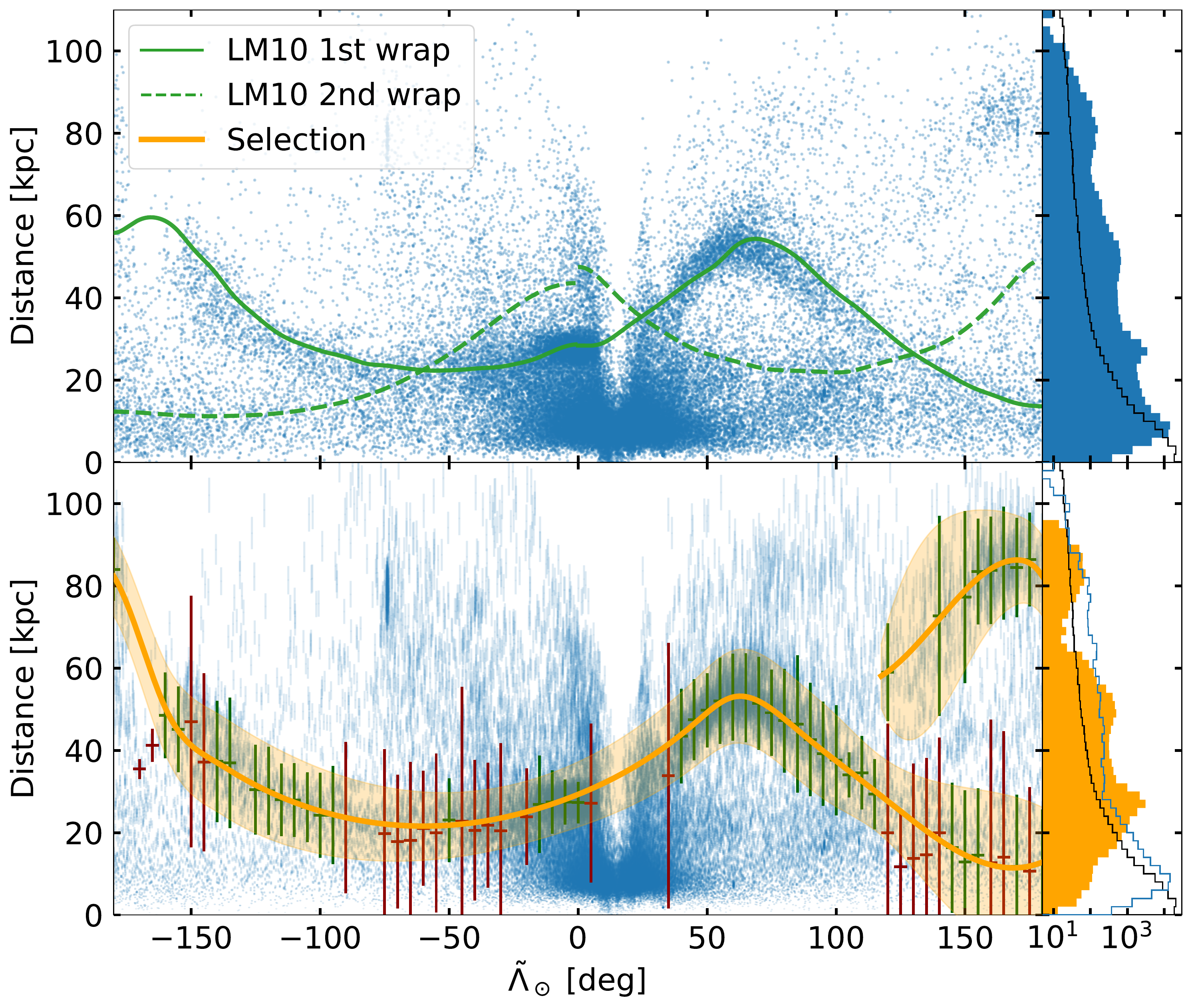}
    \caption{Same as Fig.~\ref{fig:LvsDist} for the Strip sample.}
    \label{fig:LvsDist_sky}
\end{figure}

\begin{figure}
    \centering
    \includegraphics[width=\linewidth]{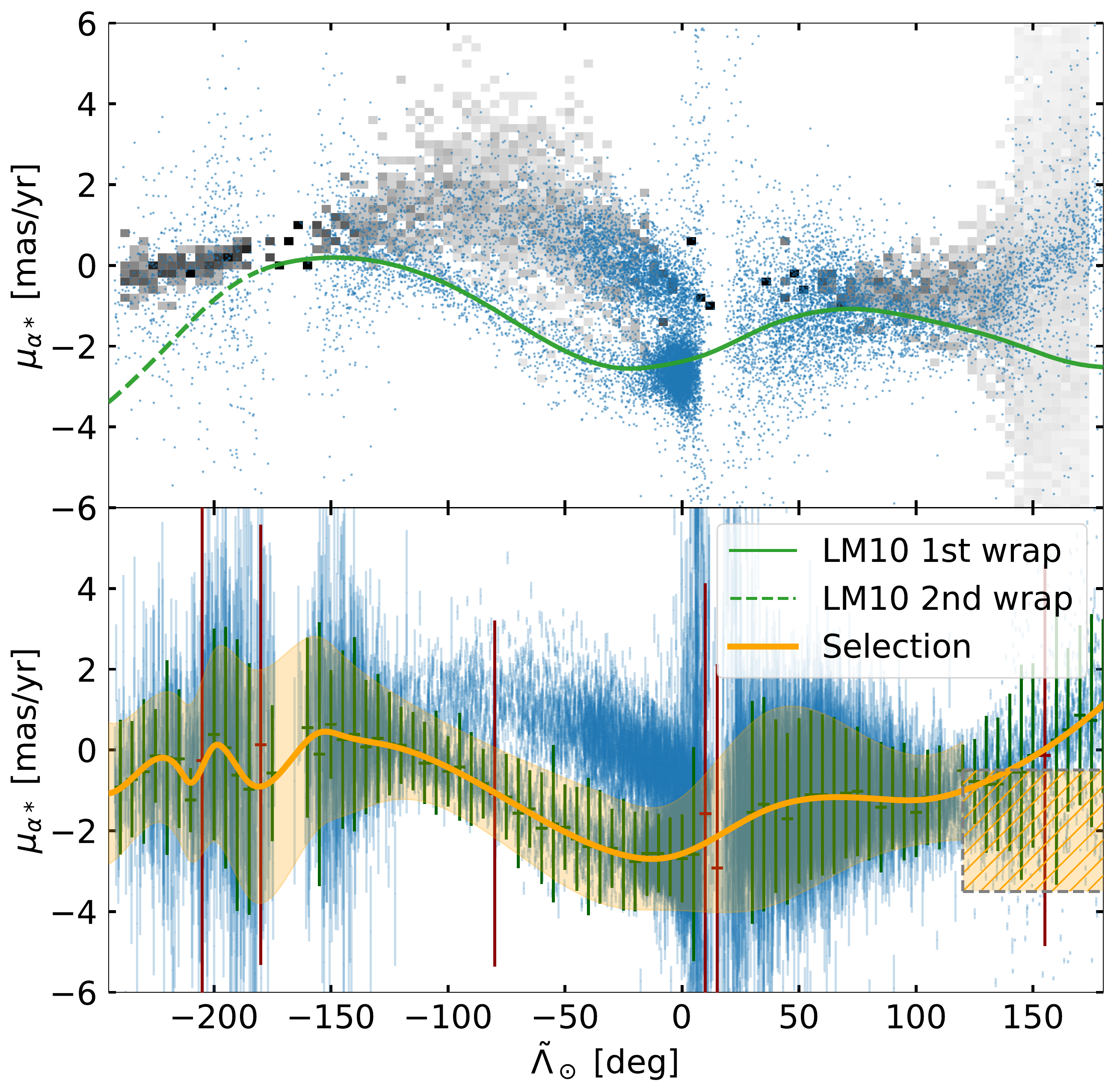}
    \caption{Same as Fig.~\ref{fig:pmra} for the Strip sample.}
    \label{fig:pmra_sky}
\end{figure}

\begin{figure}
    \centering
    \includegraphics[width=\linewidth]{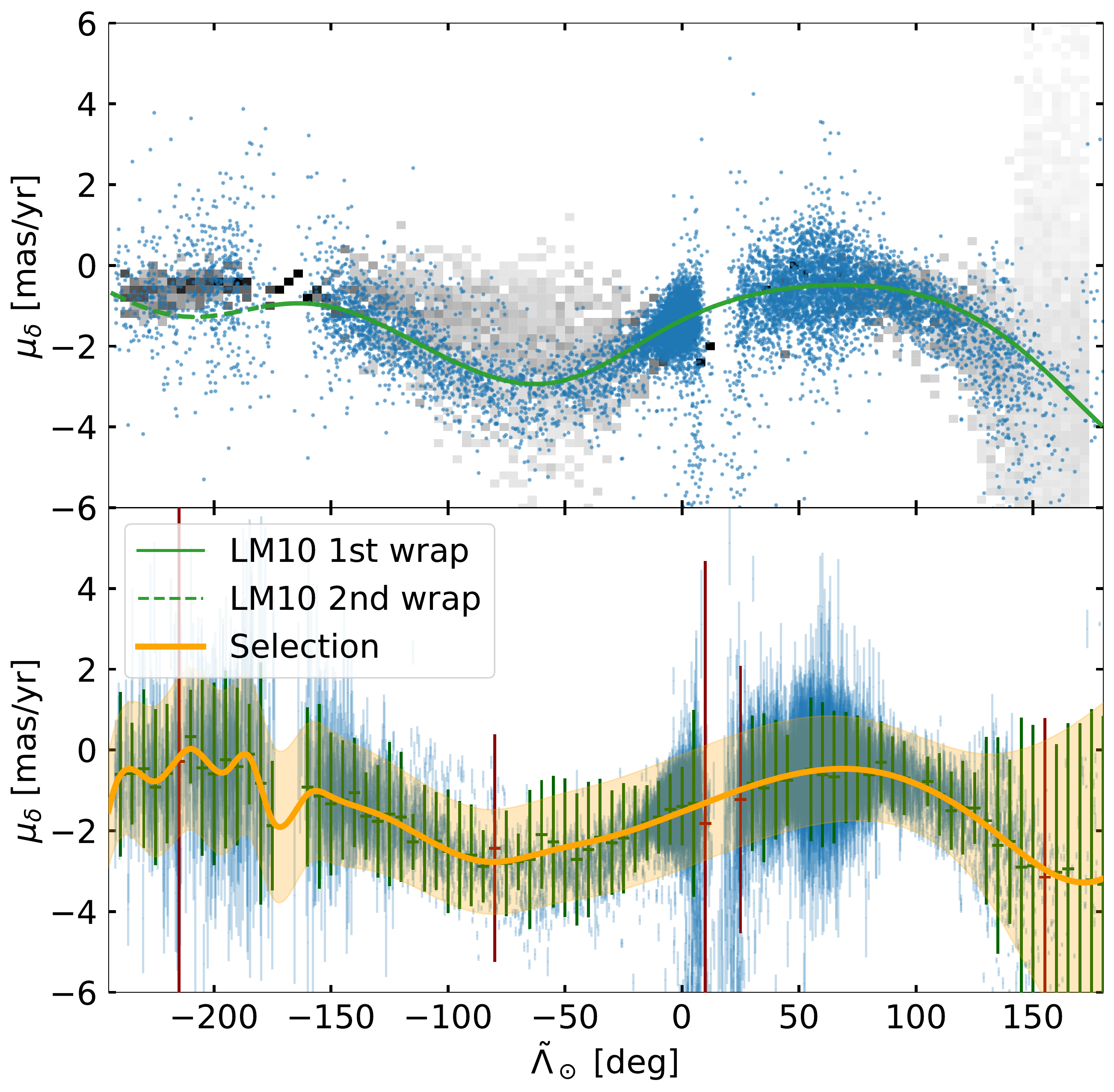}
    \caption{Same as Fig.~\ref{fig:pmdec} for the Strip sample.}
    \label{fig:pmdec_sky}
\end{figure}

Here we show the plots of each state of the filtering process starting from a Strip in the sky around the Sgr orbital plane (see Sect.~\ref{sec:initial_sample}). First, we apply the TGM to the heliocentric distance (Fig.~\ref{fig:LvsDist_sky}) then on the proper motion in right ascension (Fig.~\ref{fig:pmra_sky}) and finally, on the proper motion in declination (Fig.~\ref{fig:pmdec_sky}). The number of stars remaining after each step can be found in Table \ref{tab:counts}.

\section{Summary plots of the nGC3 sample}\label{app:nGC3_summary}

In this Appendix we discuss the summary plots of the nGC3 sample, similar to what we have done for the Strip sample in Sect.~\ref{sec:properties}.

\begin{figure}
    \centering
    \includegraphics[width=\linewidth]{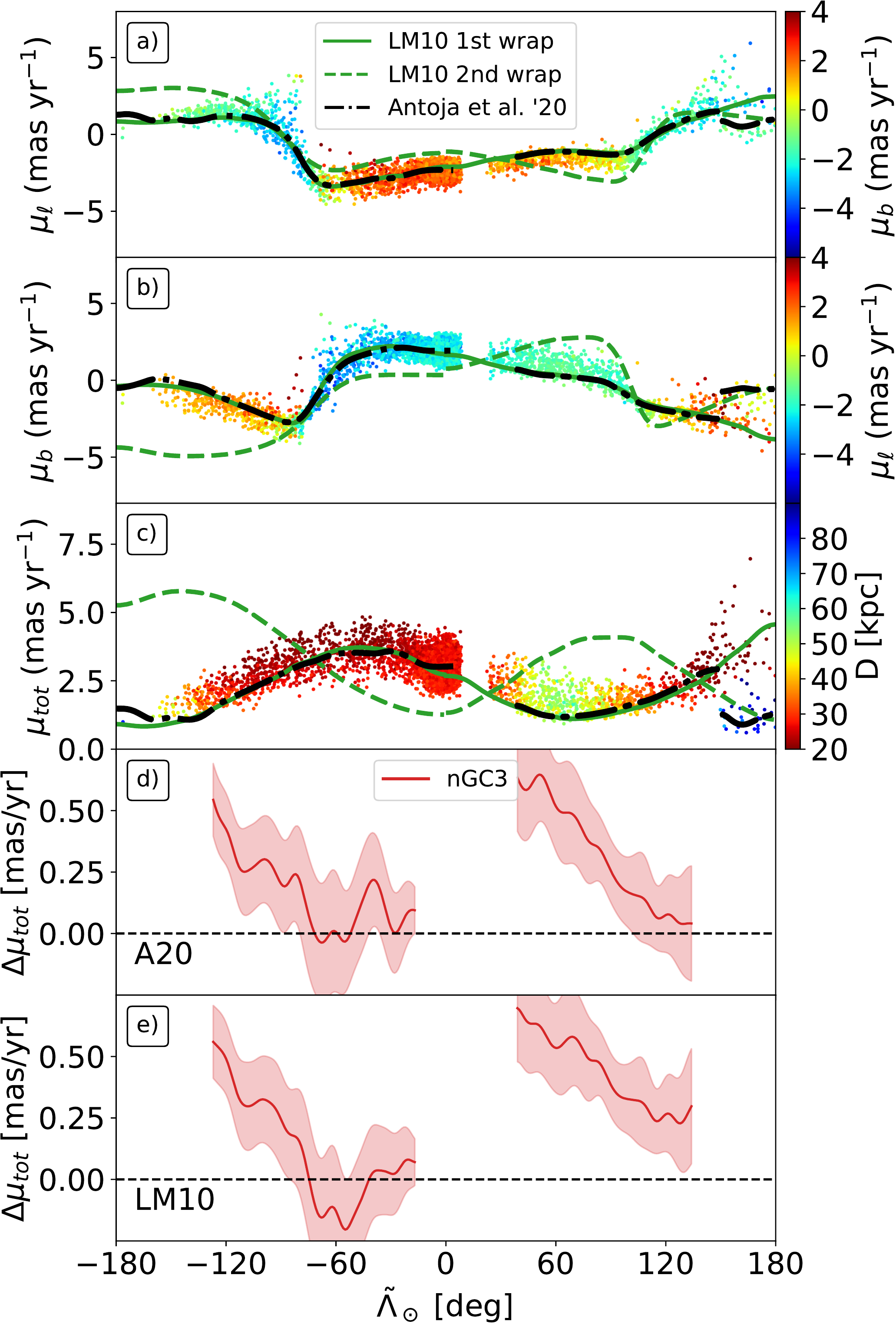}
    \caption{Same as Fig.~\ref{fig:ProperMotion} for the nGC3 sample.}
    \label{fig:ProperMotion_ngc3}
\end{figure}

Figure~\ref{fig:ProperMotion_ngc3} contains the tracks in proper motion, equivalent to Fig.~\ref{fig:ProperMotion}. With panels (a) to (c) it becomes obvious that, in the leading arm, the stars in the nGC3 sample are always on one side of the median predicted by LM10 and the observed by \citetalias{Antoja2020}. This is confirmed with the residuals, shown in panel (d), where we plot the difference between the median of the nGC3 sample and \citetalias{Antoja2020} (computed in the same way), along with the 1$\sigma$ intervals. This figure shows that even though the overall trends are similar, the discrepancy between \citetalias{Antoja2020} and the nGC3 sample exceeds 3$\sigma$ in the leading arm, becoming smaller as $\L$ increases. The difference between the nGC3 and the Strip sample in that region is also considerable, at the level of 2$\sigma$ throughout the range 0<$\L$<100$^\circ$. 
When compared to the model, additionally to the already mention bias in the leading arm (related to the lack of stars with small total proper motion) we  also observe a large discrepancy in proper motions at $\L\sim$40$^\circ$, as well as significant residuals in the trailing arm around $\L\sim$-120$^\circ$. All together, we suspect that the use of the PCM method has introduced some bias. Indeed, when we apply nGC3 algorithm to the LM10 particles we observe a similar behaviour in the residuals, thus confirming that the nGC3 sample is kinematically biased.

Regarding the trend in Galactocentric distance, we apply the same method described in Sect.~\ref{sec:GC_dist} to this sample and obtain the following apocentres for the tidal debris:

\begin{itemize}
    \item Trailing: $\L=-188^\circ \pm 2\fdg5$ at $D_{GC}= (93.22 \pm2.39)$~kpc
    \item Leading: $\L=66^\circ \pm 2\fdg5$ at $D_{GC}= (47.34 \pm0.58)$~kpc.
\end{itemize}

The values we obtain for the nGC3 and Strip samples are compatible with one another, but we note that the angular positions in the trailing apocentre differ by one bin (i.e. 5$^\circ$). Given that the nGC3 sample has few stars in that region, we expect the value inferred from the Strip to be more robust.

We have also calculated the total perpendicular velocity around M54, like we did in Sect.~\ref{sec:tan_vel} for the Strip sample, resulting in a $V_\perp$ of 274.18$\pm$0.02 \kms{}. The (small) difference between both samples is most likely caused by the larger fraction of contamination present in the Strip sample. 

\begin{figure}
    \centering
    \includegraphics[width=\linewidth]{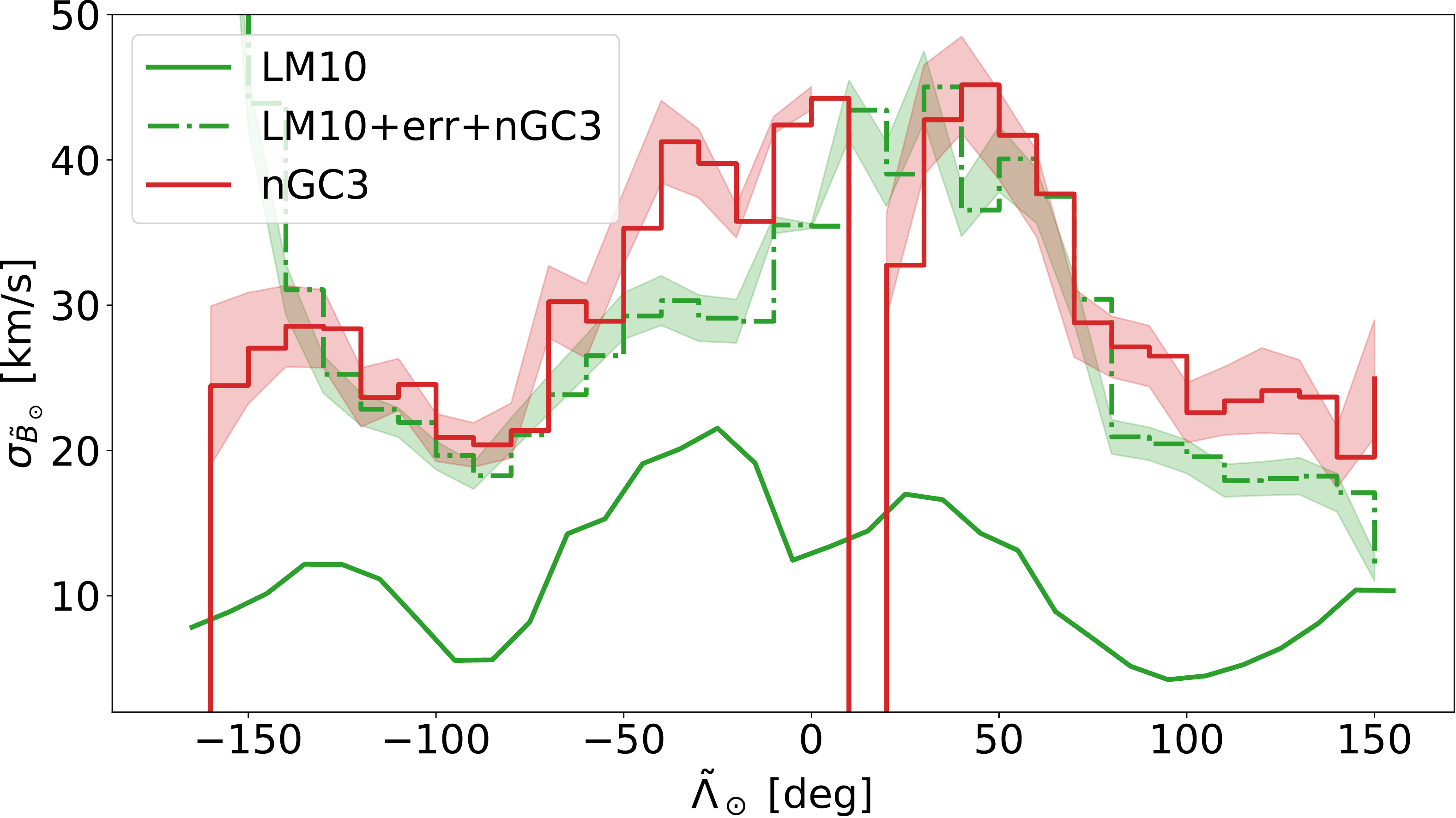}
    \caption{Same as Fig.~\ref{fig:vel_disp} but for the nGC3 sample.}
    \label{fig:vel_disp_ngc3}
\end{figure}

\begin{figure}
    \centering
\includegraphics[width=\linewidth]{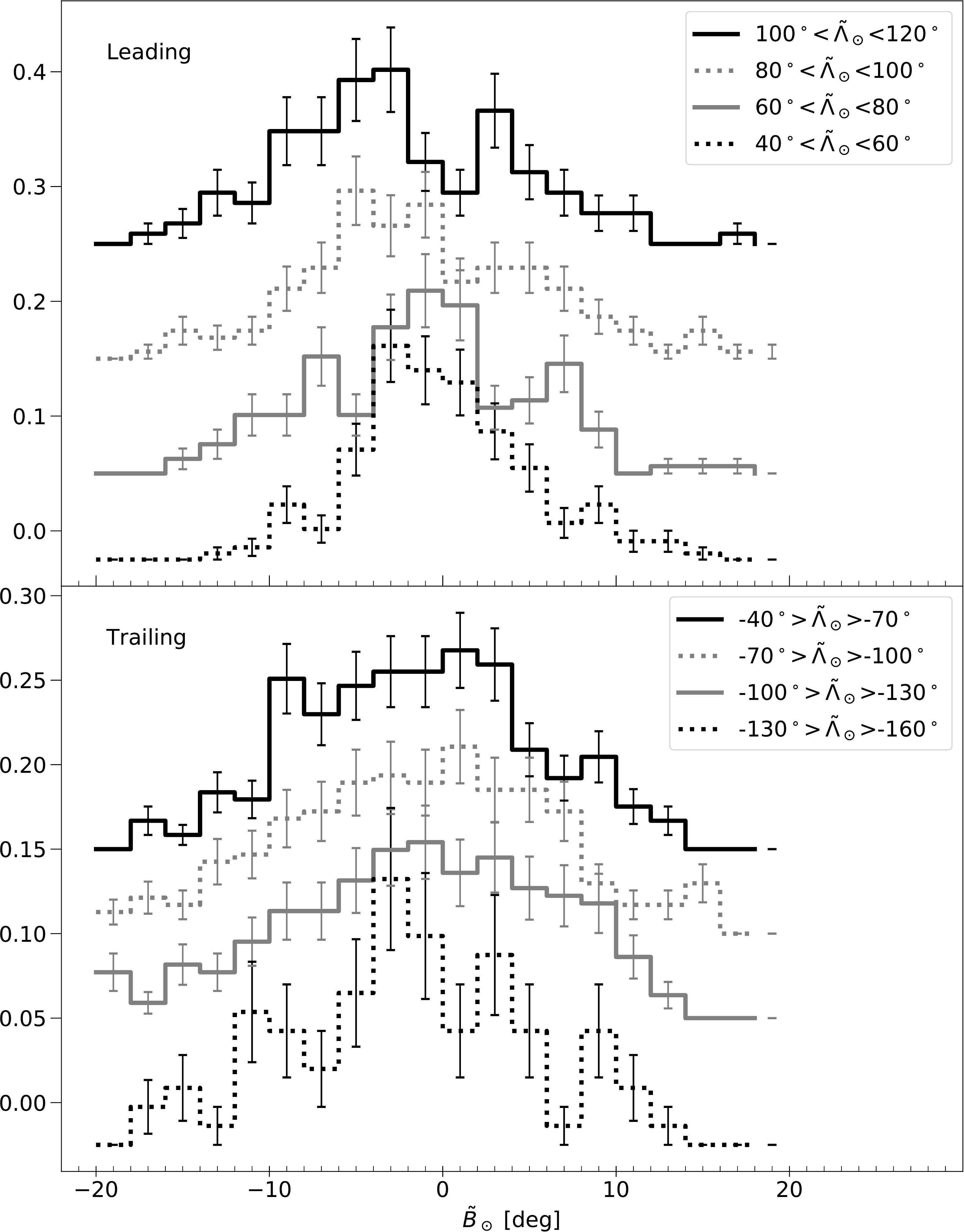}
    \caption{Same as Fig.~\ref{fig:bifu_strip} for the nGC3 sample.}
    \label{fig:bifu_ngc3}
\end{figure}

Regarding the metallicity gradients, the difference is also small. In the case of the nGC3 sample, the result is (-2.3$\pm$0.7)\,$\times 10^{-3}$\,dex\,deg$^{-1}$ in the leading arm and (1.2$\pm$0.6)\,$\times 10^{-3}$\,dex\,deg$^{-1}$ in the trailing. The velocity dispersion profile, in contrast to the Strip sample, in this case is dominated by the initial selection method: the PCM nGC3 technique. This is illustrated in Fig~\ref{fig:vel_disp_ngc3} where we show the profile derived from the nGC3 sample (red) along with the prediction of LM10 (solid green) and the result of adding realistic errors in proper motion and applying the PCM to the particles of the model (dotted-dashed line). Finally, regarding the bifurcation discussed in Sect.~\ref{sec:bifu}, it is also evident in the nGC3 sample as can be seen in Fig.~\ref{fig:bifu_ngc3}.

\section{Samples and interpolators}\label{app:interpolators}
Table~\ref{tab:sample} contains our final sample after the trimming process detailed in Sect.~\ref{sec:methods}. Along with the \begin{tt}source\_id\end{tt} and apparent magnitude in the \textit{G} band, which can be used to identify the star, we also provide the astrometry in three different Celestial coordinates: ICRS, Galactic and Sagittarius (as defined in Sect.~\ref{sec:initial_sample}). Additionally, we include the quantities that have been calculated in this work: distance and velocity along ($v_{\L}$) and across ($v_{\tilde{B}_\odot}$) the orbit. We have also added the type of star -- either RRab, RRc or RRd -- according to the SOS team, the PS1 classification and the VC table. Finally, we add two codes that can be used to filter the sources based on, respectively, their source catalogue (SOS, PS1 and/or VC) and our classification (nGC3 and/or Strip). In case of the latter, for instance, the first bit corresponds to the nGC3 sample, being 1 when the star is contained in that sample and 0 otherwise. Similarly for the Strip sample, which is represented by the second bit. As an example, a source with the code \emph{'01'} on the last column (see caption of Table~\ref{tab:sample}) has been found on the Strip sample but not on nGC3.

We also provide a series of interpolators that can be used to obtain the values of the heliocentric distance, proper motion in each reference frame (ICRS, Galactic or Sagittarius) and tangential velocities ($v_{\L}$ and $v_{\tilde{B}_\odot}$). The interpolators are given as a table containing the median of the corresponding quantity in bins of $\L$ and as a pickled \textit{Python} function (a stream of bytes that represent said object). In the case of the later, we include instructions for loading them into \textit{Python}. All this information, and more, can be found at \href{https://services.fqa.ub.edu/sagittarius}{https://services.fqa.ub.edu/sagittarius}.

\newpage
\onecolumn
\begin{landscape}

\begin{table*}\label{tab:sample}
\caption{First ten rows of the sample of RR Lyrae stars selected with our method as probable members of the Sgr stream and dwarf. For each star, we provide source\_id (column 1), \textit{G} apparent magnitude (column 2), distance (column 3) and its error (column 4) along with sky positions and proper motions, with the associated uncertainties, in three different reference systems (columns 5 to 22): ICRS, Galactic and Sgr in that order. We also provide in columns 23 to 26 the tangential velocities along and across the stream with its errors. Then, columns 27 to 29 contain the type of RR Lyrae according to, respectively, the SOS team, the PS1 classification or the VC table. Finally, column 30 contains a 3-bit code to specify in which source catalogue the source can be found: 1st bit, SOS table; 2nd bit, PS1; 3rd bit, VC. Similarly, the last column is a 2-bit code that can be used to separate between the nGC3 sample (first bit) and the Strip sample (second bit). The full table is available at the \href{http://cdsarc.u-strasbg.fr/viz-bin/cat/J/A+A/638/A104}{CDS}.}
    \footnotesize
    \setlength\tabcolsep{3pt}
    {
\begin{center}
\begin{tabular}{cccccccccc}
\hline\hline
source\_id &  G & D &  $\sigma_D$ & ra &       dec &     
$\mu_{\alpha*}$&  $\sigma_{\mu_{\alpha*}}$ &     $\mu_{\delta*}$ &  $\sigma_{\mu_{\delta}}$ \\
&  [mag]  &  [kpc] & [kpc]& [$^\circ$] & [$^\circ$] &
[mas\,yr$^{-1}$] &  [mas\,yr$^{-1}$]  & [mas\,yr$^{-1}$]  &  [mas\,yr$^{-1}$] \\
\hline
  177571127944832 &        18.176125 &  29.74 &     1.25 &  46.055586 &  0.974464 & -0.00100 &    0.337000 & -1.456000 &     0.294000 \\
  288243845193088 &        18.433370 &  34.04 &     1.44 &  44.322907 &  0.785228 &  0.44944 &    0.420931 & -1.940657 &     0.344875 \\
  782027645388032 &        17.942139 &  26.85 &     1.15 &  46.699115 &  2.303871 & -0.27400 &    0.293000 & -1.969000 &     0.339000 \\
 1035533795140608 &        17.633774 &  22.33 &     0.94 &  46.817211 &  3.026294 & -0.24200 &    0.243000 & -2.612000 &     0.234000 \\
 1407379179248512 &        17.636488 &  24.05 &     1.03 &  42.887455 &  1.805318 &  0.00500 &    0.221000 & -2.736000 &     0.205000 \\
 1742622850845696 &        18.143019 &  29.20 &     1.25 &  45.233647 &  2.930012 &  0.38200 &    0.361000 & -1.910000 &     0.353000 \\
 1889068351365504 &        18.687496 &  37.34 &     1.59 &  45.880614 &  3.615917 & -0.75300 &    0.424000 & -1.982000 &     0.388000 \\
 2018600269987200 &        18.124815 &  28.20 &     1.18 &  44.924684 &  3.642131 &  0.07200 &    0.299000 & -2.260000 &     0.265000 \\
 2057461133775616 &        18.324300 &  31.47 &     1.32 &  44.046648 &  3.646809 &  0.19800 &    0.334000 & -1.339000 &     0.337000 \\
 2345017784412032 &        17.838280 &  24.52 &     1.04 &  47.234054 &  2.984734 &  0.25200 &    0.228000 & -2.590000 &     0.233000 \\
\hline
\hline
\end{tabular}
\end{center}

\begin{center}
\begin{tabular}{cccccccccccc}
\hline\hline
l &          b &       $\mu_{l*}$&  $\sigma_{\mu_{l*}}$ & $\mu_{b}$&  $\sigma_{\mu_{b}}$ & $\L$ &  $\tilde{B}_\odot$ & $\mu_{\L*}$&  $\sigma_{\mu_{\L*}}$ & $\mu_{\tilde{B}_\odot}$&  $\sigma_{\mu_{\tilde{B}_\odot}}$\\

[$^\circ$] & [$^\circ$] &     
[mas\,yr$^{-1}$] &  [mas\,yr$^{-1}$]  & [mas\,yr$^{-1}$]  &  [mas\,yr$^{-1}$] & [$^\circ$] & [$^\circ$] &     
[mas\,yr$^{-1}$] &  [mas\,yr$^{-1}$]  & [mas\,yr$^{-1}$]  &  [mas\,yr$^{-1}$]\\
\hline
 177.010893 & -47.466387 &  1.048785 &   0.315392 & -1.009944 &   0.317069 & -118.292238 & -5.411846 &  0.725345 &   0.326881 & -1.262462 &   0.305211 \\
 175.393271 & -48.834901 &  1.687999 &   0.382442 & -1.057737 &   0.387117 & -116.690091 & -4.711547 &  0.581370 &   0.405304 & -1.905297 &   0.363113 \\
 176.272049 & -46.078912 &  1.215095 &   0.208588 & -1.573398 &   0.396562 & -119.515437 & -4.577431 &  1.213980 &   0.377142 & -1.574258 &   0.241937 \\
 175.653504 & -45.487516 &  1.681556 &   0.238517 & -2.013325 &   0.238568 & -119.977084 & -4.008319 &  1.503414 &   0.240825 & -2.149617 &   0.236238 \\
 172.740556 & -49.094217 &  1.864541 &   0.194024 & -2.002301 &   0.230696 & -115.956967 & -3.109966 &  1.367416 &   0.232648 & -2.369788 &   0.191679 \\
 174.126892 & -46.666029 &  1.601207 &   0.334025 & -1.109126 &   0.378626 & -118.553571 & -3.306850 &  0.619289 &   0.377788 & -1.846755 &   0.334972 \\
 174.109970 & -45.723944 &  0.835326 &   0.395506 & -1.948734 &   0.417007 & -119.456350 & -3.032725 &  1.637522 &   0.424904 & -1.346794 &   0.387010 \\
 173.088715 & -46.362979 &  1.599826 &   0.261816 & -1.597917 &   0.301792 & -118.640657 & -2.535687 &  1.062825 &   0.308080 & -1.995792 &   0.254387 \\
 172.147907 & -46.955375 &  1.050140 &   0.335371 & -0.854008 &   0.335636 & -117.882781 & -2.094757 &  0.496883 &   0.334750 & -1.259060 &   0.336255 \\
 176.111968 & -45.220658 &  2.024245 &   0.194627 & -1.635248 &   0.261522 & -120.319169 & -4.250378 &  1.061385 &   0.256281 & -2.375935 &   0.201477 \\

\hline
\hline
\end{tabular}
\end{center}

\begin{center}
\begin{tabular}{cccccccccccc}
\hline\hline
$v_{\L}$ &  $\sigma_{v_{\L}}$ &    $v_{\tilde{B}_\odot}$ &  $\sigma_{v_{\tilde{B}_\odot}}$& SOS & PS1 & VC &  Catalogue &  Sample \\

[\kms{}] & [\kms{}] & [\kms{}] & [\kms{}] & & & & & & \\
\hline
 156.23 &      46.55 &  63.64 &      43.11 &      - &     RRab &     - &               010 &       01 \\
 147.25 &      65.70 & -65.71 &      58.66 &     RRab &     RRab &    RRab &              111 &       01 \\
 208.87 &      48.83 &  41.18 &      30.84 &      - &     RRab &    RRab &               011 &      11 \\
 213.67 &      27.04 &  13.94 &      25.02 &      - &      RRc &     - &               010 &      11 \\
 209.06 &      27.99 & -28.47 &      21.89 &      - &     RRab &    RRab &               011 &      11 \\
 139.78 &      52.64 & -14.05 &      46.37 &      - &     RRab &    RRab &               011 &      11 \\
 344.24 &      76.64 &   3.08 &      68.51 &      - &     RRab &    RRab &               011 &       01 \\
 196.16 &      41.99 & -25.29 &      34.02 &      RRc &      RRc &     RRc &              111 &      11 \\
 127.96 &      50.22 &  53.63 &      50.21 &      - &      RRc &     - &               010 &       01 \\
 177.98 &      30.74 & -34.68 &      23.47 &      - &     RRab &    RRab &               011 &      11 \\

\hline
\hline
\end{tabular}    
\end{center}

}
\end{table*}
\end{landscape}

\end{appendix}

\end{document}